\documentclass[noinfoline]{imsart}

\usepackage{amsthm,amsmath}
\RequirePackage{natbib}
\RequirePackage{hyperref}

\usepackage{amssymb}
\usepackage{mathptmx}
\usepackage{graphicx,psfrag,epsf}
\usepackage{enumerate}
\usepackage{multirow}
\usepackage{bm}
\usepackage{color}
\usepackage{floatrow}
\usepackage{units}
\usepackage{verbatim}
\usepackage{epstopdf}
\usepackage{booktabs,caption,fixltx2e}
\usepackage{rotating}
\usepackage{multirow}
\usepackage{physics}
\usepackage{afterpage}
\usepackage[section]{placeins} 
\usepackage{longtable}
\usepackage{array}
\usepackage[table]{xcolor}
\usepackage{wrapfig}
\usepackage{float}
\usepackage{colortbl}
\usepackage{pdflscape}
\usepackage{tabu}
\usepackage{threeparttable}
\usepackage{threeparttablex}
\usepackage[normalem]{ulem}
\usepackage{makecell}
\usepackage{adjustbox}
\usepackage[font=footnotesize]{caption}
\usepackage[linesnumbered,ruled]{algorithm2e}
\usepackage{nameref}
\usepackage{fancyhdr}
\usepackage{enumitem}
\usepackage{listings}

\newtheorem{theorem}{Theorem}

\theoremstyle{definition}

\makeatletter
\renewcommand*{\@fnsymbol}[1]{\ensuremath{\ifcase#1\or a\or b\or *
\else\@ctrerr\fi}}
\makeatother

\begin{document}

\definecolor{light-gray}{gray}{0.95}
\lstset{columns=fullflexible, basicstyle=\ttfamily,
    backgroundcolor=\color{white},xleftmargin=0.5cm,frame=lr,framesep=8pt,framerule=0pt}

\begin{frontmatter}

\title{Pseudo Bayesian Mixed Models under Informative Sampling}
\runtitle{Pseudo Bayesian Mixed Models}


\author{\fnms{Terrance D.} \snm{Savitsky}\thanksref{t1}\ead[label=e1]{savitsky.terrance@bls.gov}}
\and
\author{\fnms{Matthew R.} \snm{Williams}\thanksref{t2,t3}\ead[label=e2]{mrwillia@nsf.gov}}

\thankstext{t1}{U.S. Bureau of Labor Statistics, Office of Survey Methods Research,
 Washington, DC, USA,
 savitsky.terrance@bls.gov
 }
\thankstext{t2}{National Center for Science and Engineering Statistics,
       National Science Foundation,
       Alexandria, VA, USA,
	mrwillia@nsf.gov
       }
\thankstext{t3}{Corresponding Author}

\runauthor{Savitsky and Williams}

\begin{abstract}
When random effects are correlated with survey sample design variables, the usual approach of employing individual survey weights (constructed to be inversely proportional to the unit survey inclusion probabilities) to form a pseudo-likelihood no longer produces asymptotically unbiased inference.  We construct a weight-exponentiated formulation for the random effects distribution that achieves approximately unbiased inference for generating hyperparameters of the random effects. We contrast our approach with frequentist methods that rely on numerical integration to reveal that only the Bayesian method achieves both unbiased estimation with respect to the sampling design distribution and consistency with respect to the population generating distribution. Our simulations and real data example for a survey of business establishments demonstrate the utility of our approach across different modeling formulations and sampling designs.  This work serves as a capstone for recent developmental efforts that combine traditional survey estimation approaches with the Bayesian modeling paradigm and provides a bridge across the two rich but disparate sub-fields.
\end{abstract}


\end{frontmatter}
\section{Introduction}
Hierarchical Bayesian models provide a flexible and powerful framework for social science and economic data, which often include nested units of analysis such as industry, geography, and individual.
Yet, social science and economic data are commonly acquired from a survey sampling procedure.  It is typically the case that the underlying survey sampling design distribution governing the procedure induces a correlation between the response variable(s) of interest and the survey sampling inclusion probabilities assigned to units in the target population from which the sample was taken. Survey sampling designs where there is a correlation between the response variable and the sampling inclusion probabilities are referred to as informative and will result in the distribution of the response variable in observed samples being different from that from the underlying population about which we seek to perform model-based inference.  Sample designs may also be informative when the inclusion probabilities for groups are correlated with the corresponding latent random effects. The resulting distribution of random effects in the sample is different from that of the population of random effects.

The current literature for Bayesian methods has partially addressed population model estimation of survey data under informative sampling designs through the use of survey sampling weights to obtain consistent estimates of fixed effects or top level global parameters. Yet the survey statistics literature suggests that parameters related to random effects, or local parameters are still potentially estimated with bias. The possibility for survey-induced bias in estimation of random effects severely limits the applicability of the full suite of Bayesian models to complex social and economic data.  

This paper proposes a Bayesian survey sample-weighted, plug-in framework for the simultaneous estimation of fixed effects and generating hyperparameters (e.g., variance) of random effects that is unbiased with the respect to the distribution over samples and asymptotically consistent with respect to the population distribution.

\subsection{Informative Sampling Designs}
Survey sampling designs that induce a correlation between the response variable of interest, on the one hand, and the survey sample inclusion probabilities, on the other hand, are deemed ``informative" and produce samples that express a different balance of information than that of the underlying population, thus estimation methods that do not incorporate sample design information lead to incorrect inferences. For example, the U.S. Bureau of Labor Statistics (BLS) administers the Job Openings and Labor Turnover Survey (JOLTS) to business establishments for the purpose of estimating labor force dynamics, such as the total number of hires, separations and job openings for area-indexed domains.  The units are business establishments and their inclusion probabilities are set to be proportional to their total employment (as obtained on a lagged basis from a census instrument).  Since the number of hires, separations and job openings for establishments are expected to be correlated to the number of employees, this survey sampling design induces informativeness, so that hiring, separations and openings would be expected to be larger in the samples than in the underlying population.

\subsection{Bayesian Models for Survey Data}
There is growing and rich literature on employment of survey sampling weights (constructed to be inversely proportional to unit inclusion probabilities) for correction of the population model estimated on the observed survey sample to produce asymptotically unbiased estimation.   Some recent papers focus on the use of Bayesian modeling for the specific purpose of producing mean and total statistics under either empirical or nonparametric likelihoods, but these methods don't allow the data analyst to specify their own population models for the purpose of parameter estimation and prediction \citep{kunihama:2014,wu:2010,si2015}. In particular, the set-up for our paper is one where the data analyst has specified a particular Bayesian hierarchical model for the population (from which the sample was taken) under which they wish to perform inference from data taken from a complex sampling design.  So, having to specify a model that is specific to the realized sample, but unrelated to the population model constructed by the data analyst does not allow them to conduct the inference they seek. \citet{2015arXiv150707050S} and \citet{2018dep} complement these efforts by employing a pseudo-posterior to allow the data analyst to estimate a population model of their choice on an informative sample taken from that population.  The pseudo-likelihood exponentiates each unit likelihood contribution by its sampling weight, which re-balances information in the observed sample to approximate that in the population.  The use of the pseudo-posterior may be situated in the more general class of approximate or composite likelihoods \citep{2009arXiv0911.5357R}.   All of the above Bayesian approaches that allow analyst specification of the underlying population generating model to be estimated on the observed informative sample \emph{only} address models with fixed or global effects, not random effects.  Yet, it is routine in Bayesian modeling to employ one or more sets of random effects under prior formulations designed to capture complex covariance structures.   Hierarchical specifications make such population models readily estimable.

\subsection{Extending the Pseudo-Posterior to Mixed Effects Models}
There are two survey designs considered in this paper: 1. Clusters or groups of units are sampled in a first stage, followed by the sampling of nested units in a second stage.  We refer to this procedure as the ``direct" sampling of clusters or groups; 2. Units are sampled in a single stage without directly sampling the clusters or groups in which they naturally nest (e.g., geography).
We refer to the case where groups used in the population model are not included in the sampling design as ``indirect" sampling of groups, since a group is included in the sample to the extent that a nested unit is directly sampled.

This paper extends the approaches of \citet{2015arXiv150707050S} and \citet{2018dep} from global-only parameters to mixed effects (global and local parameter) models by exponentiating both the data likelihood contributions \emph{and} the group-indexed random effects prior distributions by sampling weights - an approach that we label, ``double-weighting" - that is multiplied, in turn, by the joint prior distribution for the population model parameters to form a joint pseudo-posterior distribution with respect to the observed data and random effects for the sampled groups.

Our augmented (by sample-weighting the prior for random effects) pseudo-posterior method introduced in the next section is motivated by a data analyst who specifies a population generating (super population) model that includes group-indexed random effects for which they desire to perform inference.   The observed data are generated under an informative sampling design such that simply estimating the population model on the observed sample will produce biased parameter estimates.  Our augmented pseudo-posterior model estimator uses survey sampling weights to perform a relatively minor adjustment to the model augmented likelihood such that parameter draws taken on the observed informative sample approximates inference with respect to the population generating distribution (and the resulting parameter estimates are consistent with respect to the population generating distribution).

We demonstrate that our pseudo-posterior formulation achieves both unbiasedness with respect to the two-stage survey sampling design distribution \emph{and} consistency with respect to the population generating (super population) model for the observed response variable and the latent cluster/group random effects under both direct and indirect sampling of groups.  Our weighted pseudo-posterior is relatively simple in construction because we co-sample the random effects and model parameters in our numerical Markov chain Monte Carlo (MCMC) scheme and marginalize out the random effects \emph{after} estimation.

The case of indirect sampling of groups is particularly important in Bayesian estimation as it is common to specify multiple random effects terms that parameterize a complex covariance structure because the random effects terms are readily estimable in an MCMC scheme.

The remainder of the paper proceeds to develop our proposed double-weighting methods for estimation of mixed effects models under both direct and indirect sampling of groups on data acquired under informative sampling in Section~\ref{doublepseudo}.   Simulation studies are conducted in Section~\ref{simulation} that compare our proposed method to the usual case of likelihood weighting under direct sampling of groups. Section~\ref{application} applies our double-weighting method to estimation of the number of hires for business establishments under employment of industry-indexed random effects in the population model where we reveal that our double-weighting approach produces more heterogeneous random effects estimates to better approximate the population from the observed sample than does the usual practice.  We offer a concluding discussion in Section~\ref{discussion}.

\section{Mixed Effects Pseudo Posterior Distribution}\label{doublepseudo}
The focus of this paper addresses sampling units naturally collected into a population of groups; for example, defined by geography.  There is typically a dependence among the response values for units within each group such that units are more similar within than between groups.  Sampling designs are typically constructed as multi-stage where the collection of groups in the population are first sampled, followed by the sequential taking of a sub-sample of units from the population of units within \emph{selected} or sampled groups.   

By contrast, an alternative set of sampling designs may proceed to draw a sample from the population of units in a \emph{single} stage such that the groups are included the sample, indirectly, when one or more of their member units are selected.  These two sampling designs - sampling groups, followed by sampling units within groups in a multi-stage sampling design, on the one hand, as compared to a single-stage sampling of units without directly sampling their member groups, on the other hand - will lead us to design two formulations for extending the pseudo-posterior distribution of \citet{2018dep}. 

The pseudo-posterior exponentiates the likelihood contributions of the \emph{single level} fixed effects model (that does not utilize random effects) by the survey weights for the observed sample units, which are inversely proportional to their probabilities of being selected into the sample $w_{i} \propto 1/\pi_{i}$:
\begin{equation}\label{eq:pseudopost}
    f^{\pi}\left(\theta\vert \mathbf{y},\tilde{\mathbf{w}}\right) \propto \left[\textcolor{red}{\mathop{\prod}_{i = 1}^{n}f\left(y_{i}\vert \theta\right)^{\tilde{w}_{i}}}\right]f\left(\theta\right)
\end{equation}
where the normalized weights sum to the sample size $\tilde{w}_{i} = w_{i}/\frac{\sum w_{i}}{n}$.  The sum of the weights directly affects the amount of posterior uncertainty estimated in the posterior distribution for $\bm{\theta}$, so normalizing it to sum to the effective sample size regulates that uncertainty. Equation \ref{eq:pseudopost} is a noisy approximation to the true (but not fully known) joint distribution of the population model $P_{\theta_{0}}(\mathbf{y})$ and the sampling process $P^{\pi}(\bm{\delta})$, where $\bm{\delta}$ denotes a vector of sample design inclusion indicators for units and groups (that are governed by $P^{\pi}$) and formally defined under the 2-stage class of sampling designs considered in this paper in the upcoming section. 

The noisy approximation for the population likelihood obtained by constructing the sample-weighted pseudo-posterior estimator for the observed (informative) sample leads to consistent estimation of population generating (super-population) parameters $\theta$ (as the sample size, $n$, grows) .  

The pseudo-posterior construction requires only a minor change to the population model specified by the data analyst on which they wish to perform inference (by weighting each unit-indexed likelihood contribution by its associated marginal sampling weight).  In particular, the data analyst may specify population distributions for $f\left(y_{i}\vert \theta\right)$ and priors $f(\bm{\theta})$; for example, if the data are count data that we work with in the sequel, they may specify a Poisson likelihood with mean, $\bm{\mu}$, for which they define a latent regression model formulation.  The data analyst is interested to perform inference for the generating parameters under the population generation and not the distribution of the observed sample.   Under informative sampling the two distributions are different and the pseudo-posterior corrects the distribution of the observed sample back to the population of interest.  

We demonstrate in the sequel, that the formulation in Equation \ref{eq:pseudopost} can be extended to multi-level models by exponentiating \emph{both} the likelihood (conditioned on the random effects) \emph{and} the prior distribution for random effects by sampling weights.

\subsection{Mixed Effects Posterior Under \emph{Direct} Sampling of Population Groups}\label{sec:direct}
Assign units, $i \in (1,\ldots,N)$, that index a population, $U$, to groups, $h \in (1,\ldots,G_{U})$, where each population group, $h$, nests $U_{h} = 1,\ldots,N_{h}$ units, such that $N = \lvert U \rvert = \mathop{\sum}_{h = 1}^{G_{U}}N_{h}$, with $N_{h} = \lvert U_{h} \rvert$.  Construct a 2-stage informative sampling design whose first stage takes a direct sample of the $G_{U}$ groups, where $\pi_{h}\in(0,1]$ denotes the marginal sample inclusion probability for group, $h \in (1,\ldots,G_{U})$.  Let $g \in (1,\ldots,G_{S})$, index the \emph{sampled} groups, where $G_{S}$ denotes the number of observed groups from the population of groups, $G_{U} \supset G_{S}$.

Our first result defines a pseudo-posterior estimator on the observed sample for our population model that includes group-indexed random effects in the case where we \emph{directly} sample groups, followed by the sampling of units nested within selected groups, in a multistage survey sampling design.  Our goal is to achieve unbiased inference for $(\theta,\phi)$ (where $\theta$ denotes fixed effects for generating population responses, $y$, and $\phi$ denotes the generating parameters of random effects, $u$, for the population), estimated on our observed sample taken under an informative survey sampling design.  
We assume that random effects $u$ are indexed by group and are independent conditional on the generating parameter $\phi$.
Multistage designs that sample groups or clusters, followed by the further sampling of nested units, are commonly used for convenience to mitigate costs of administration where in-person interviews are required and also in the case where a sampling frame of end-stage units is constructed after sampling groups in the first stage. The second stage of the survey sampling design takes a sample from the $N_{g}$ (second stage) units $\forall g \in S_{g}$, where $S_{g} \subset U_{g}$.  The second stage units are sampled with conditional inclusion probabilities, $\pi_{\ell\mid g} \in (0,1]$ for $\ell = 1,\ldots,N_{g}$, conditioned on inclusion of group, $g\in (1,\ldots,G_{S})$. Let $j \in (1,\ldots,n_{g})$ index the sampled or observed second stage units linked to or nested within sampled group, $g \in (1,\ldots,G_{S})$.  Denote the marginal group survey sampling weight, $w_{g} \propto 1/\pi_{g}$ for $\pi_{g} \in (0,1]$.  Denote the marginal unit survey sampling weight, $w_{g j} \propto 1/\pi_{g j}$, for $\pi_{g j} \in (0,1]$, the joint inclusion probability for unit, $j$, nested in group, $g$, both selected into the sample. The group marginal inclusion probabilities and conditional unit inclusion probabilities under our 2-stage survey sampling design are governed by distribution, $P^{\pi}$.

\begin{theorem}\label{th:direct}
Under a proper prior specification, $f(\bm{\theta})f(\bm{\phi})$, the following pseudo-posterior estimator achieves approximately unbiased inference for super-population (population generating) model, $f(\bm{\theta},\bm{\phi}|\mathbf{y})$, with respect to the distribution governing the taking of samples from an underlying finite population, $P^{\pi}$,
\begin{equation}\label{eq:sampdirmodel}
f^{\pi}\left(\theta, \phi \vert  \mathbf{y} \right) \propto
\left[\mathop{\int}_{\mathbf{u} \in \mathcal{U}} \left\{ 
\textcolor{blue}{\mathop{\prod}_{g\in S}}
\left(
\textcolor{red}{\mathop{\prod}_{j \in S_{g}} f\left(y_{g j}\mid u_{g}, \theta \right)^{w_{g j}}}
\right)
\textcolor{blue}{f\left(u_{g}\mid \phi \right)^{w_g}} \right\} d\mathbf{u} \right]f(\theta)f(\phi).
\end{equation}
where $f^{\pi}(\cdot)$ denotes a sampling-weighted pseudo-distribution, $j \in S_{g}$ denotes the subset of units, $j \in (1,\ldots,n_{g} = \lvert S_{g}\rvert)$, linked to group, $g\in (1,\ldots,G_{S})$.   Parameters, $(\bm{\theta},\bm{\phi})$, index the super-population model posterior distribution, $f(\bm{\theta},\bm{\phi}|\mathbf{y})$, that is the target for estimation. The integral for the vector $\mathbf{u} = (u_{1},\ldots,u_{n_{g}})$ is taken over its support, $\mathcal{U}$, for each component, $u_{g} \in \mathbf{u}$.
\end{theorem}
We employ a pseudo-likelihood for the first level of the model for sampled observations $y_{gj}$ within sampled clusters $g$ by exponentiating by the sample weight $w_{gj}$. This provides a noisy approximation to the first stage likelihood. For the second level model (or prior) for the random effects $u_{g}$, we exponentiate this distribution by its corresponding sampling weights $w_{g}$. This provides a noisy approximation to the population distribution of random effects. Both approximations are needed because the distributions of both the responses and the random effects in the sample can differ substantially from those in the corresponding population due to the informative sampling design at both the cluster $g$ and the within cluster $j|g$ stages, where the latter notation denotes the sampling of unit $j$ conditioned on / within sampled group $g$.

Under our augmented (by weighting the prior of the group-indexed random effects) pseudo-likelihood of Equation~\ref{eq:sampindirmodel}, $f\left(y_{g j}\mid u_{g}, \theta \right)$ and $f\left(u_{g}\mid \phi \right)$ are not restricted; for example, we select a Poisson distribution for the observed data likelihood, $f\left(y_{g j}\mid u_{g}, \theta \right)$, for our simulation study and application in the sequel.  Similarly, the form of the distribution for the random effects prior distribution is not restricted under our construction, though it is most commonly defined as Gaussian under a GLM specification.  Replacing the single Gaussian with a mixture of Gaussian distributions would also fit our set-up.   Our approach also readily incorporates additional levels of random effects with no conceptual changes.

\begin{proof}
We first construct the complete joint model for the finite population, $U$, as if the random effects, $(u_{h})$, were directly observed,
\begin{equation}\label{eq:popmodel}
f_{U}\left(\theta, \phi \vert  \mathbf{y},\mathbf{u} \right) \propto
\left[\mathop{\prod}_{h = 1}^{G_{U}}
\left(\mathop{\prod}_{\ell= 1}^{N_{h}} f\left(y_{h \ell}\mid u_{h}, \theta \right)\right)
f\left(u_{h}\mid \phi \right)  \right]f(\theta)f(\phi).
\end{equation}

Under a complex sampling design, we specify random sample inclusion indicators for groups, $\delta_h$, with marginal probabilities $\pi_{h} = P(\delta_h= 1)$ for $h \in (1,\ldots,G_{U})$, governed by $P^{\pi}$.  We further specify random sample inclusion indicator, $\delta_{\ell \mid h} = (\delta_{\ell} \mid \delta_h = 1) \in \{0,1\}$, with probability $\pi_{\ell\mid h} = P(\delta_{\ell \mid h}= 1)$, for unit $\ell \in (1,\ldots,N_{h})$, conditioned on the inclusion of group, $h$, such that the indicator for the joint sampling of unit $\ell$ nested within group $h$ is denoted as $\delta_{h\ell} = \delta_{\ell \mid h} \times \delta_{h}$, with the associated marginal inclusion probability, $\pi_{h\ell} = P(\delta_{h\ell}=1)$.

The taking of an observed sample is governed by the survey sampling distribution, $P^{\pi}$ (as contrasted with $P_{\theta,\phi}$, the population generation distribution for $(y,u)$).  The pseudo-likelihood with respect to the joint distribution, $(P^{\pi},P_{\theta,\phi})$, is then constructed by exponentiating components of the likelihood in the population such that the \emph{expected value} of the survey sample pseudo log-likelihood function with respect to $P^{\pi}$ equals that of the log-likelihood for the entire population (and thus the score functions also match in expectation).  Let $\ell_{U}\left(\mathbf{y}, \mathbf{u} \vert \theta, \phi \right) \equiv \log f_{U}\left(\theta, \phi \vert  \mathbf{y},\mathbf{u} \right)$ denote the population model log-likelihood. Applying this approach to the log-likelihood of the joint model, above, leads to the following pseudo- likelihood formulation:
\begin{align}
\ell ^{\pi}_{U}\left(\mathbf{y}, \mathbf{u} \vert \theta, \phi \right) &\propto
\mathop{\sum}_{h=1}^{G_{U}} \left(\mathop{\sum}_{\ell = 1}^{N_{h}} \left(\frac{\delta_{\ell\mid h}}{\pi_{\ell\mid h}}\right)\left(\frac{\delta_h}{\pi_h}\right) \ell \left(y_{h \ell}\mid u_{h}, \theta \right)\right)
+
 \left(\frac{\delta_h}{\pi_h}\right) \ell \left(u_{h}\mid \phi \right)\\
 & = \mathop{\sum}_{h=1}^{G_{U}} \left(\mathop{\sum}_{\ell = 1}^{N_{h}} \left(\frac{\delta_{h \ell}}{\pi_{h \ell}}\right) \ell \left(y_{h \ell}\mid u_{h}, \theta \right)\right)
+
 \left(\frac{\delta_h}{\pi_h}\right) \ell \left(u_{h}\mid \phi \right)\label{estimator}
\end{align}
where $P^{\pi}$ governs all possible samples, $(\delta_{h},\delta_{\ell\mid h})_{\ell\in U_{h}, h=1,\ldots,G_{U}}$, taken from population, $U$. Let joint group-unit inclusion indicator, $\delta_{h \ell} = \delta_{h}\times\delta_{\ell\mid h}$ with $\pi_{h \ell} = P(\delta_{h \ell}= 1) = P(\delta_{h}=1,\delta_{\ell\mid h} = 1)$.  For each \emph{observed} sample $\ell ^{\pi}_{U}\left(\mathbf{y}, \mathbf{u} \vert \theta, \phi \right) = \ell ^{\pi}_{S}\left(\mathbf{y}, \mathbf{u} \vert \theta, \phi \right)$ where

\begin{equation}\label{estimatorsample}
\ell ^{\pi}_{S}\left(\mathbf{y}, \mathbf{u} \vert \theta, \phi \right) =
\mathop{\sum}_{g=1}^{G_{S}} \left(\mathop{\sum}_{j \in S_{g}} w_{g j} \ell \left(y_{g j}\mid u_{g}, \theta \right)\right)
+
 w_g \ell \left(u_{g}\mid \phi \right)
\end{equation}
and $w_{g j} \propto \pi_{g,j}^{-1}$ and $w_g \propto \pi_{g}^{-1}$. The expectation of our estimator in Equation~\ref{estimator} is unbiased with respect to $P^{\pi}$,
\begin{align}
\mathbb{E}^{\pi}\left[\ell ^{\pi}_{U}\left(\mathbf{y}, \mathbf{u} \vert \theta, \phi \right)\bigg\vert P_{\theta,\phi}\right] &\equiv \\
\mathbb{E}^{\pi}\left[\ell ^{\pi}_{U}\left(\mathbf{y}, \mathbf{u} \vert \theta, \phi \right)\right]  &= \ell_{U}\left(\mathbf{y}, \mathbf{u} \vert \theta, \phi \right)\label{unbiased},
\end{align}
where the expectation, $\mathbb{E}^{\pi}(\cdot)$, is taken with respect to the survey sampling distribution, $P^{\pi}$, that governs the survey sampling inclusion indicators, $\{\delta_{h \ell}, \delta_h\}$, conditional on the data $\{\mathbf{y}, \mathbf{u}\}$ generated by $P_{\theta,\phi}$.  The final equality in Equation~\ref{unbiased} is achieved since $\mathbb{E}^{\pi}(\delta_{h \ell}) = \pi_{h \ell}$ and $\mathbb{E}^{\pi}(\delta_{h}) = \pi_{h}$.


Thus, we use the following sampling-weighted model approximation to the complete population model of Equation~\ref{eq:popmodel}:
\begin{equation}\label{eq:sampcompletemodel}
f^{\pi}\left(\theta, \phi \vert  \mathbf{y}, \mathbf{u}\right) \propto \left[\mathop{\prod}_{g \in S}\left(\mathop{\prod}_{j \in S_{g}} f\left(y_{g j}\mid u_{g}, \theta \right)^{w_{g j}}\right)
f\left(u_{g}\mid \phi \right)^{w_g} \right]f(\theta)f(\phi).
\end{equation}
We can then construct a sampling-weighted version of the observed model:
\begin{equation}\label{eq:sampobsmodel}
f^{\pi}\left(\theta, \phi \vert  \mathbf{y} \right) \propto
\left[\mathop{\int}_{\mathbf{u} \in \mathcal{U}} \left\{ \mathop{\prod}_{g \in S}
\left(\mathop{\prod}_{j \in S_{g}} f\left(y_{g j}\mid u_{g}, \theta \right)^{w_{g j}}\right)
f\left(u_{g}\mid \phi \right)^{w_g} \right\} d \mathbf{u} \right]f(\theta)f(\phi).
\end{equation}

The walk from Equation~\ref{eq:sampcompletemodel} to Equation~\ref{eq:sampobsmodel} is possible because we co-estimate the $(\mathbf{u})$ with $(\theta,\phi)$ and then perform the integration step to marginalize over the $(\mathbf{u})$ \emph{after} estimation.
\end{proof}

Theorem~\ref{th:direct} requires the exponentiation of the prior contributions for the sampled random effects, $(u_{g})$, by a sampling weight, $w_{g} \propto 1/\pi_{g}$ in order to achieve approximately unbiased inference for $\phi$; it is not enough to exponentiate each data likelihood contribution, $f\left(y_{g j}\mid u_{g}, \theta \right)$, by a unit (marginal) sampling weight, $w_{g j}$.   This formulation is generally specified for \emph{any} population generating model, $P_{\theta,\phi}$.  Our result may be readily generalized to survey sampling designs of more than two stages where each collection of later stage groups are nested in earlier stage groups (such as households of units nested within geographic PSUs).

The proposed method under direct sampling of Equation~\ref{eq:sampdirmodel} is categorized as a plug-in estimator that exponentiates the likelihood contributions for nested units by the unit-level marginal sampling weights \emph{and}, in turn, exponentiates the prior distribution for cluster-indexed random effects by the cluster (or PSU) marginal sampling weights.
Samples from the joint pseudo-posterior distribution over parameters are interpreted as samples from the underlying (latent) population generating model since the augmented (by the weighted random effects prior distribution) pseudo-likelihood estimated on the observed sample provides a noisy approximation for the population generating likelihood.

Although the augmented pseudo-likelihood is unbiased with respect to the distribution over samples, it will not, generally, produce correct uncertainty quantification.  In particular, the credibility intervals will be too optimistic relative to valid frequentist confidence intervals because the plug-in method is not fully Bayesian in that it doesn't model uncertainty in the sampling weights, themselves.
Although the employment of random effects captures dependence among nested units, the warping and scaling induced by the sampling weights will result in failure of Bartlett's second identity such that the asymptotic hyperparameter covariance matrix for our plug-in mixed effects model will not be equal to the sandwich form of the asymptotic covariance matrix for the MLE.  The result of the lack of equality is that the model credibility intervals, without adjustment, will not contract on valid frequentist confidence intervals.  

A recent work of \citet{leonnovelo2021fully} jointly models the unit level marginal sampling weights and the response variable and includes group-indexed random effects parameters in their joint model.   They demonstrate correct uncertainty quantification because the asymptotic covariance matrix of their fully Bayesian model (that also co-models the sampling weights) is equal to that for the MLE.  Their method specifies an exact likelihood for the observed sample that is complicated and requires a closed-form solution for an integral that restricts the class of models that may be considered.   This approach requires a different model formulation than that specified for the population and of interest to the data analyst.

By contrast, our plug-in augmented pseudo-posterior distribution requires only minor change to the underlying population model specified by the data analyst and may be easily adapted to complicated population models.

Correct uncertainty quantification for $(\bm{\theta},\bm{\phi})$ may be achieved by using the method of \citet{2018arXiv180711796W} to perform a post-processing of the MCMC parameter draws that replaces the pseudo-posterior covariance with the sandwich form of the MLE.  This paper, by contrast, focuses on providing unbiased point estimation for mixed effects models as an extension of \citet{2015arXiv150707050S} because \citet{2018arXiv180711796W} may be readily applied, post sampling.

Our pseudo-likelihood in Equation~\ref{estimator} is jointly conditioned on $(\mathbf{y},\mathbf{u})$, such that the random weights, $\left(\delta_{h \ell}/\pi_{h \ell}\right)$, are specified in linear summations.  This linear combination of weights times log-likelihoods ensures (design) unbiasedness with respect to $P^{\pi}$ because the weight term is separable from the population likelihood term.   We may jointly condition on $(\mathbf{y},\mathbf{u})$ in our Bayesian set-up because we co-sample $(\mathbf{u},\theta,\phi)$, numerically, in our MCMC such that the integration step over $\mathbf{u}$ is applied \emph{after} co-estimation.  In other words, we accomplish estimation in our MCMC by sampling $\mathbf{u}$ jointly with $(\theta,\phi)$ on each MCMC iteration and then ignoring $\mathbf{u}$ to perform marginal inferences on $\theta$ and $\phi$, which is a common approach with Bayesian hierarchical models.  By contrast, \citet{pfeffmix:1998}, \citet{rh:1998}, and others \citep[for example see][]{kim2017statistical} specify the following integrated likelihood under frequentist estimation for an observed sample where units are nested within groups,
\begin{equation}
\ell^{\pi}(\theta,\phi) = \mathop{\sum}_{g=1}^{G_{S}} w_{g} \ell_{i}^{\pi}(\theta,\phi),
\end{equation}
for $\ell_{i}^{\pi}(\theta,\phi) = \log L_{i}^{\pi}(\theta,\phi)$ for,
\begin{equation}\label{eq:frequnitlike}
L_{i}^{\pi}(\theta,\phi) = \mathop{\int}_{u_{g} \in \mathcal{U}}\exp\left[\mathop{\sum}_{j \in S_{g}} w_{j\mid g}\ell \left(y_{g j}\mid u_{g}, \theta \right)\right]f\left(u_{g}\mid \phi \right)du_{g},
\end{equation}
which will \emph{not}, generally, be unbiased with respect to the distribution governing the taking of samples for the population likelihood because the unit level conditional weights, $(w_{j\mid g})_{j}$, are nested inside an exponential function (such that replacing $w_{j\mid g}$ with $\delta_{\ell\mid h}/\pi_{\ell \mid h}$ inside the exponential and summing over the population groups and nested units will not produce separable sampling design terms that each integrate to $1$ with respect to $P^{\pi}$, conditioned on the generated population) \citep{yi:2016}.  The non-linear specification in Equation~\ref{eq:frequnitlike} results from an estimation procedure that integrates out $\mathbf{u}$ \emph{before} pseudo-maximum likelihood point estimation of $(\theta,\phi)$.

This design biasedness (with respect to $P^{\pi}$) is remedied for pseudo-maximum likelihood estimation by \citet{yi:2016} with their alternative formulation,
\begin{align}
\ell^{\pi}(\theta,\phi) &=\mathop{\sum}_{g=1}^{G_{S}} w_{g} \mathop{\sum}_{j < k; j,k\in S_{g}} w_{j,k\mid g} \ell_{gj,k}(\theta,\phi)\label{composite}\\
\ell_{gj,k}(\theta,\phi) &= \log \left\{\mathop{\int}_{u_{g} \in \mathcal{U}}f(y_{g j}\mid u_{g},\theta)f(y_{g k}\mid u_{g},\theta)f(u_{g}\mid\phi)du_{g} \right\},
\end{align}
where $w_{j,k\mid g} \propto 1/\pi_{j,k\mid g}$ denotes the joint inclusion probability for units $(j,k)$, both nested in group, $g$, conditioned on the inclusion of group, $g$, in the observed sample.  Equation~\ref{composite} specifies an integration over $u_{g}$ for \emph{each} $f(y_{g j}\mid u_{g},\theta)f(y_{g k}\mid u_{g},\theta)$ pair, which allows the design weights to enter in a linear construction outside of each integral.  This set-up establishes linearity for inclusion of design weights, resulting in unbiasedness with respect to the distribution governing the taking of samples for computation of the pseudo-maximum likelihood estimate, though under the requirement that pairwise unit sampling weights be published to the data analyst or estimated by them. 

Yet, the marginalization of the random effects \emph{before} applying the group weight, $w_{g}$, fails to fully correct for the prior distribution for $u_{g}$.  We show in the sequel that $\bm{\phi}$ is estimated with bias by \citet{yi:2016} due to this integration of the random effects being performed on the unweighted prior of $u_g$. Our method, by contrast, weights the prior for $u_{g}$ and performs the integration of $u_{g}$ numerically in our MCMC (after sampling $u_{g}$).

\subsection{Mixed Effects Posterior Under \emph{Indirect} Sampling of Population Groups}\label{sec:indirect}
Bayesian model specifications commonly employ group-level random effects (often for multiple simultaneous groupings) to parameterize a complex marginal covariance structure.   Those groups are often not \emph{directly} sampled by the survey sampling design.  We, next, demonstrate that weighting the prior contributions for the group-indexed random effects is \emph{still} required, even when the groups are not directly sampled, in order to achieve unbiased inference for the generating parameters of the random effects, $\phi$. Again, as throughout, we assume the group-indexed random effects are conditionally independent given generating parameter $\phi$. We focus our result on a simple, single-stage sampling design, that may be readily generalized, where we reveal that the group-indexed survey sampling weights are constructed from unit marginal inclusion probabilities. Constructing sampled group weights from those of member units appeals to intuition because groups are included in the observed sample only if any member unit is selected under our single-stage survey sampling design.

Suppose the same population set-up as for Theorem~\ref{th:direct}, with population units, $\ell \in U_{h}$, linked to groups, $h \in (1,\ldots,G_{U})$, where each unit, $(h,\ell)$, maps to $i \in (1,\ldots,N)$.  We now construct a \emph{single} stage sampling design that directly samples each ($h,\ell$) unit with marginal inclusion probability, $\pi_{h \ell}$, governed by $P^{\pi}$.  Group, $g \in G_{S}$, is \emph{indirectly} sampled based on whether there is \emph{any} linked unit, ($g j$), observed in the sample.

\begin{theorem}\label{th:indirect}
The following pseudo-posterior estimator achieves approximately unbiased inference with respect to $P^{\pi}$,

\begin{equation}\label{eq:sampindirmodel}
\begin{split}
f^{\pi}\left(\theta, \phi \vert  \mathbf{y} \right) \propto
&\left[\mathop{\int}_{\mathbf{u} \in \mathcal{U}}  \left\{
\textcolor{blue}{ \mathop{\prod}_{g\in S}}
\left(
\textcolor{red}{\mathop{\prod}_{j \in S_{g}} f\left(y_{g j}\mid u_{g}, \theta \right)^{w_{g j}}}
\right)\right.\right.\\
&\left.\left.\vphantom{
            \mathop{\int}_{u_{g} \in \mathcal{U}}\left(\mathop{\prod}_{j \in S_{g}} f\left(y_{g j}\mid u_{g}, \theta \right)^{w_{g j}}\right)
    }
    \textcolor{blue}{ f\left(u_{g}\mid \phi \right)^{w_g = \frac{1}{N_{g}}\displaystyle\mathop{\sum}_{j\in S_{g}}w_{g j}}}
     \right\} d \mathbf{u} \right]f(\theta)f(\phi),
\end{split}
\end{equation}

where $w_{g j} \propto 1/\pi_{gj}$.
\end{theorem}
\begin{proof}
We proceed as in Theorem~\ref{th:direct} by supposing the population $U$ of units and associated group-indexed random effects, $(u_{h})$, were fully observed.  We first construct the likelihood for the fully observed population.
\begin{align}
f_{U}\left(\theta, \phi \vert  \mathbf{y},\mathbf{u} \right) &\propto
\left[\mathop{\prod}_{h = 1}^{G_{U}}
\left(\mathop{\prod}_{\ell= 1}^{N_{h}} f\left(y_{h \ell}\mid u_{h}, \theta \right)\right)
f\left(u_{h}\mid \phi \right)  \right]f(\theta)f(\phi)\label{eq:popunitindirect}\\
&=\left[\mathop{\prod}_{h = 1}^{G_{U}}
\mathop{\prod}_{\ell= 1}^{N_{h}} \left\{f\left(y_{h \ell}\mid u_{h}, \theta \right)
f\left(u_{h}\mid \phi \right)^{\frac{1}{N_{h}}} \right\} \right]f(\theta)f(\phi)\label{eq:popindirect}.
\end{align}
We proceed to formulate the pseudo-likelihood for all possible random samples taken from $U$, $f^{\pi}_{U}(\cdot)$, governed jointly by $(P^{\pi},P_{\theta,\phi})$, from which we render the pseudo-likelihood for any sample, $f^{\pi}(\cdot)$, which is constructed to be unbiased with respect to the distribution governing the taking of samples for the population model of Equation~\ref{eq:popindirect} under $P^{\pi}$,
\begin{align}\label{eq:popindmodel}
f^{\pi}_{U}\left(\theta, \phi \vert  \mathbf{y}, \mathbf{u}\right) &\propto \left[\mathop{\prod}_{h=1}^{G_{U}}\mathop{\prod}_{\ell \in U_{h}} \left\{f\left(y_{h \ell}\mid u_{h}, \theta \right)f\left(u_{h}\mid \phi \right)^{\frac{1}{N_{h}}}\right\}^{\frac{\delta_{h \ell}}{\pi_{h \ell}}} \right]f(\theta)f(\phi)\\
&=\left[\mathop{\prod}_{h=1}^{G_{U}}f\left(u_{h}\mid \phi \right)^{\frac{1}{N_{h}}\mathop{\sum}_{\ell\in U_{h}}\frac{\delta_{h \ell}}{\pi_{h \ell}}}\mathop{\prod}_{\ell \in U_{h}} f\left(y_{h \ell}\mid u_{\ell}, \theta \right)^{\frac{\delta_{h \ell}}{\pi_{h \ell}}}\right]f(\theta)f(\phi).
\end{align}
This pseudo-posterior reduces to the following expression for the observed sample,
\begin{equation}
f^{\pi}\left(\theta, \phi \vert  \mathbf{y}, \mathbf{u}\right) \propto \left[\mathop{\prod}_{g=1}^{G_{S}}f\left(u_{g}\mid \phi \right)^{\frac{1}{N_{g}}\mathop{\sum}_{j\in S_{g}}w_{g j}}\mathop{\prod}_{j \in S_{g}} f\left(y_{g j}\mid u_{j}, \theta \right)^{w_{g j}}\right]f(\theta)f(\phi),
\end{equation}
where $\pi_{g j} = P(\delta_{g j} = 1)$ (under $P^{\pi}$), $w_{g j} \propto 1/\pi_{g,j}$ and $N_{g}$ denotes the number of units in the population linked to observed group, $g \in(1,\ldots,G_{S})$ observed in the sample.  We set $w_{g} := 1/N_{g}\times \mathop{\sum}_{j\in S_{g}}w_{g j}$ and the result is achieved.
\end{proof}
This result derives from eliciting group-indexed weights from unit inclusion probabilities for units linked to the groups.  While the resulting pseudo-posterior estimators look very similar across the two theorems, the sampling designs are very different from one another in that groups are not directly sampled in this latter case, which is revealed in their differing formulations for $w_{g}$.  

The averaging of unit weights formulation for $w_{g}$  naturally arises under the derivation of Equation~\ref{eq:popindirect} when sampling units, rather than groups under a model that utilizes group-indexed random effects to capture within group dependence that naturally arises among units in the population. 
Exponentiating the augmented pseudo-likelihood of Equation~\ref{eq:popunitindirect} by survey variables anticipates the integration of the random effects to produce an observed data pseudo-likelihood.  We may intuit this result by interpreting this form for $w_{g}$ proportional to the average importance of units nested in group each group, $g$.  It bears mention that in the indirect sampling case, there is no probability of group selection defined for a single stage design.

In practice, it is not common for the data analyst to know the population group sizes, $(N_{g})$, for the groups, $g\in (1,\ldots,G_{S})$ observed in the sample, so one estimates an $\hat{N}_{g}$ to replace $N_{g}$ in Equation~\ref{eq:sampindirmodel}.  Under a single stage sampling design where the groups are indirectly sampled through inclusions of nested units into the observed sample, we assume that we only have availability of the marginal unit inclusion sampling weights, $(w_{g j})$.  The group population size, $N_{g}$, needed for the sum-weights method of Equation~\ref{eq:sampindirmodel}, may be estimated by $\hat{N}_{g} = \mathop{\sum}_{j = 1}^{n_{g}}w_{j\mid g}$.  To approximate $w_{j\mid g}$, we first utilize the sum-probabilities result to estimate, $\hat{w}_{g} = 1/\hat{\tilde{\pi}}_{g}$, and proceed to extract $(w_{j\mid g})$ from $w_{gj} \approx w_{g}w_{j\mid g}$.  If we invert the resultant group-indexed weight, $w_{g}= 1/N_{g}\times\mathop{\sum}_{j \in S_{g}}w_{g j}$, for the case where groups are not directly sampled, we may view the inverse of the group $g$ weight, $\tilde{\pi}_{g} = 1/w_{g}$, as a ``pseudo" group inclusion probability, since we don't directly sample groups.  One may envision other formulations for the pseudo group inclusion probabilities, $\tilde{\pi}_{g}$, that we may, in turn, invert to formulate alternative group-indexed survey sampling weights, $(w_{g})$.  Please see Appendix \ref{sec:pseudoprobs} where we develop other methods, in addition to sum-weights, for computing $\tilde{\pi}_{g}$.

In application, we normalize the by-group, survey sampling weights, $\displaystyle (w_{g})_g = 1,\ldots,G_{S}$, to sum to the number of observed groups in the sample, $G_{S}$, and normalize unit weights, $(w_{gj})_{j = 1,\ldots,n_{g}}$ to sum to the overall sample size, $n$.  These normalizations regulate uncertainty quantification for posterior estimation of $(u_{g})$ and global parameters, $(\phi,\theta)$ by encoding an effective number of observed groups and units.  So, we normalize them to sum to the number of groups and units observed in the sample to regulate the estimated pseudo-posterior variance of $(\phi,\theta)$.  (In practice, these normalizations often produce somewhat optimistic credibility intervals due to dependence induced by the survey sampling design. \citet{2018arXiv180711796W} provide an algorithm that adjusts pseudo-posterior draws to incorporate this dependence).

We refer to our proposed procedure for weight exponentiating both the data likelihood contributions and the prior distributions of the $(u_{g})$ as ``double-weighting", as mentioned in the introduction, to be contrasted with the usual approach of ``single-weighting" of \citet{2018dep} developed for models with global effects parameters.

\section{Simulation study}\label{simulation}
Our simulation study in the sequel focuses on a count data response rather than the usual continuous response, both because count data are the most common data type for the employment data collected by BLS and because our Bayesian construction is readily estimable under any response data type.

We generate a count data response variable, $y$, for a population of size, $N = 5000$ units, where the logarithm of the generating mean parameter, $\mu$ is constructed to depend on a size predictor, $x_{2}$, in both fixed and random effects terms; in this way, we construct both fixed and random effects to be informative, since our proportion-to-size survey sampling design sets unit inclusion probabilities to be proportional to $x_{2}$.   We generate a population of responses using,
\begin{align}\label{eq:margmodel}
y_{i} &\sim \mathcal{P}\left(\mu_{i}\right) \nonumber\\
\log\mu_{i} &= \beta_{0} + x_{1i}\beta_{1} + x_{2i}\beta_{2} + \left[1,x_{2i}\right]\bm{\gamma}_{h\{i\}},
\end{align}
where $\mathcal{P}(\cdot)$ denotes the Poisson distribution, $x_{1i} \sim \mathcal{N}\left(0,1\right)$ is the inferential predictor of interest to the data analyst and $x_{2i} \sim \mathcal{E}\left(1/2.5\right)$ (where $\mathcal{E}(\cdot)$ denotes the Exponential distribution) is the size variable, which is generated from a skewed distribution to reflect real-world survey data, particularly for employment counts.  
The expression, $h\{i\}$, denotes the group $h \in (1,\ldots,G_{U})$ linked to unit $i\in (1,\ldots,N)$.  We generate $\displaystyle\mathop{\bm{\gamma}_{h}}^{2\times 1} \sim \mathcal{N}_{2}\left(\mathbf{0},\mbox{diag}(\bm{\sigma})\times \mathop{R}^{2\times 2} \times \mbox{diag}(\bm{\sigma})\right)$, where $\bm{\sigma} = (1.0,0.5)^{'}$.  We set $R = \mathbb{I}_{2}$, where $\mathbb{I}_{2}$ denotes the identity matrix of size $2$.  Finally, we set $\bm{\beta} = \left(\beta_{0}, \beta_{1},\beta_{2}\right)^{'} = (0, 1.0,0.5)$, where we choose the coefficient of $x_{2}$ to be lower than that for $x_{1}$ to be moderately informative, which is conservative.

The allocation of units, $i = 1,\ldots,N$ to groups, $h = 1,\ldots,G_{U}$ is performed by sorting the units, $i$, based on size variable, $x_{2}$.  This allocation procedure constructs sized-based groups that accord well with survey designs that define groups as geographic regions, for convenience, where there is expected more homogeneity within groups than between groups.

The population size for each group, $N_{h}$, is fixed under direct sampling of groups; for example, $N_{h} = 4$ in the case of $G_{U} = 1250$, which produces $N = 5000$ units, so the number of population units per group is constructed as $(4, 10, 25, 50, 100)$ for population group sizes, $G_{U} = (1250,500,200,100,50)$, respectively.

Although the population response $y$ is generated with $\mu = f(x_1,x_2)$, we estimate the marginal model for the \emph{population}, $\mu = f(x_1)$ to which we will compare estimated results on samples taken from the population to assess bias and mean-squared error (MSE). We use $x_{2}$ in the generation of the population values for $y$ because the survey sampling inclusion probabilities are set proportionally to $x_{2}$, which instantiates the informativeness of the sampling design.   In practice, however, the analyst does not have access to $x_{2}$ for the population units or, more generally, to all the information used to construct the survey sampling distribution that sets inclusion probabilities for all units in the population.  The marginal estimation model under exclusion of size variable, $x_{2}$, is specified as:

\begin{align}
\label{eq:estmodel}
y_{i} &\sim \mathcal{P}\left(\mu_{i}\right) \nonumber\\
\log\mu_{i} &= \beta_{0} + x_{1i}\beta_{1} + u_{h\{i\}}\\
u_{h} & \sim \mathcal{N}\left(0,\sigma_{u}^{2}\right) \nonumber
\end{align}
 where now $u_{h}$ is an intercept random effect, $h = 1,\ldots,G_{U}$.  
 
 Our goal is to estimate the global parameters, $(\beta_{0},\beta_{1},\sigma_{u}^{2})$, from informative samples of size, $n = 500$, taken from the population (of size, $N = 5000$).  We utilize the following simulation algorithm:
\begin{enumerate}
    \item Each Monte Carlo iteration of our simulator (that we run for $B = 300$ iterations) generates the population $(y_{i},x_{1i},x_{2i})_{i=1}^{N}$ from Equation~\ref{eq:margmodel} on which we estimate the marginal population model of Equation~\ref{eq:estmodel} to determine the population true values for $(\mu_{i},\sigma_{u}^{2})$.  
\item Our simulation study focuses on the       direct sampling of groups, followed by a     sub-sampling of units within the            selected groups. We use a                   proportion-to-size design to directly       sample from the $G_{U}$ groups in the       first stage, where the group inclusion      probabilities, $\pi_{h} \propto      
    \frac{1}{N_{h}}\mathop{\sum}_{i\in U_{h}}x_{2i}$.   
    
    We draw a sample of groups in the first stage and observe $G_{S} < G_{U}$ groups. In particular, fixed sample of total size $n = 500$ is taken where the number of groups sampled, $G_{S} = n/(Nf) \times G_{U}$.
    
\item The second stage size-based sampling      of units is accomplished with inclusion     probabilities, $\pi_{\ell\mid g} \propto     x_{2\ell}$ for $\ell \in
    (1,\ldots,N_{g})$. 
    
    We perform a further sub-sampling of $f\%$ of population units in the selected $G_{S}$ groups.
    
\item Estimation is performed for  $(\beta_{0},\beta_{1},\sigma_{u}^{2})$ from the observed sample of $n = 500$ under three alternatives:
    \begin{enumerate}
    \item Single-weighting, where we solely exponentiate the likelihood contributions for $(y_{g j})$ by sampling weights, $(w_{g j} \propto 1/\pi_{g j})$ (and don't weight the prior for the random effects, $(u_{g})$);
    \item Double-weighting, where we exponentiate \emph{both} the likelihood for the $(y_{g j})$ by sampling weights, $(w_{g j})$, and also exponentiate the prior distribution for $u_{g}$ by weight, $w_{g} \propto 1/\pi_{g}$ (for each of $g = 1,\ldots,G_{S}$). We compute the marginal unit weights used in both single- and double-weighting as $w_{g j} \propto 1/\pi_{g j}$, where $\pi_{g j}$ is the marginal inclusion probability, formulated as, $\pi_{g j} = \pi_{g}\pi_{j\mid g}$ for $ j = 1,\ldots,n_{g}$ for each group, $g \in 1,\ldots, G_{S}$ in the case of direct sampling of groups.  
    \item SRS, where in the case of direct sampling of groups, we take a simple random (equal probability) sample of groups in a first stage, followed by a simple random sample of units within selected groups.
    We take the SRS sample from the same population as is used to take the two-stage, pps informative sample. 
    The inclusion of model estimation under (a non-informative) SRS is intended to serve as a gold standard against which we may judge the bias and MSE performance of single- and double-weighting under informative sampling.
    \end{enumerate}
\end{enumerate}

We use Stan \citep{stan:2015} to estimate the double-weighted mixed effects model of Equation~\ref{eq:sampcompletemodel} for the specific case of the Poisson likelihood that we use in our simulations and application that next follows.  We fully specify our Stan probability model for the Poisson likelihood under double-weighting in Appendix~\ref{sec:stanscript}.  In particular, we specify a multivariate Gaussian joint prior distribution for the $K\times 1$, $\bm{\beta}$ coefficients with a vector of standard deviation parameters, $\sigma_{\beta}$ drawn from a truncated Cauchy prior.  The associated correlation matrix for the multivariate Gaussian prior for $\bm{\beta}$ is drawn from a prior distribution that is uniform over the space of $K \times K$ correlation matrices.  The prior for the standard deviation parameter of the random effects, $\sigma_{u}$, is also specified as a truncated Cauchy distribution.   These prior distributions are designed as weakly informative by placing large probability mass over a large support region, while expressing a mode to promote regularity and a stable posterior geometry that is readily explored under Stan's Hamiltonian Monte Carlo scheme. 
The single-weighting case is achieved as a special / simplified case of the double-weighting model.

Please see Appendix \ref{sec:simstudy} for results of a second simulation study under the \emph{indirect} sampling of groups.

\newpage

\subsection{Informative Random Effects Under Direct Sampling of Groups}
\begin{figure}
\centering
\includegraphics[width = 0.80\textwidth,
		page = 1,clip = true, trim = 0.0in 0.0in 0in 0.in]{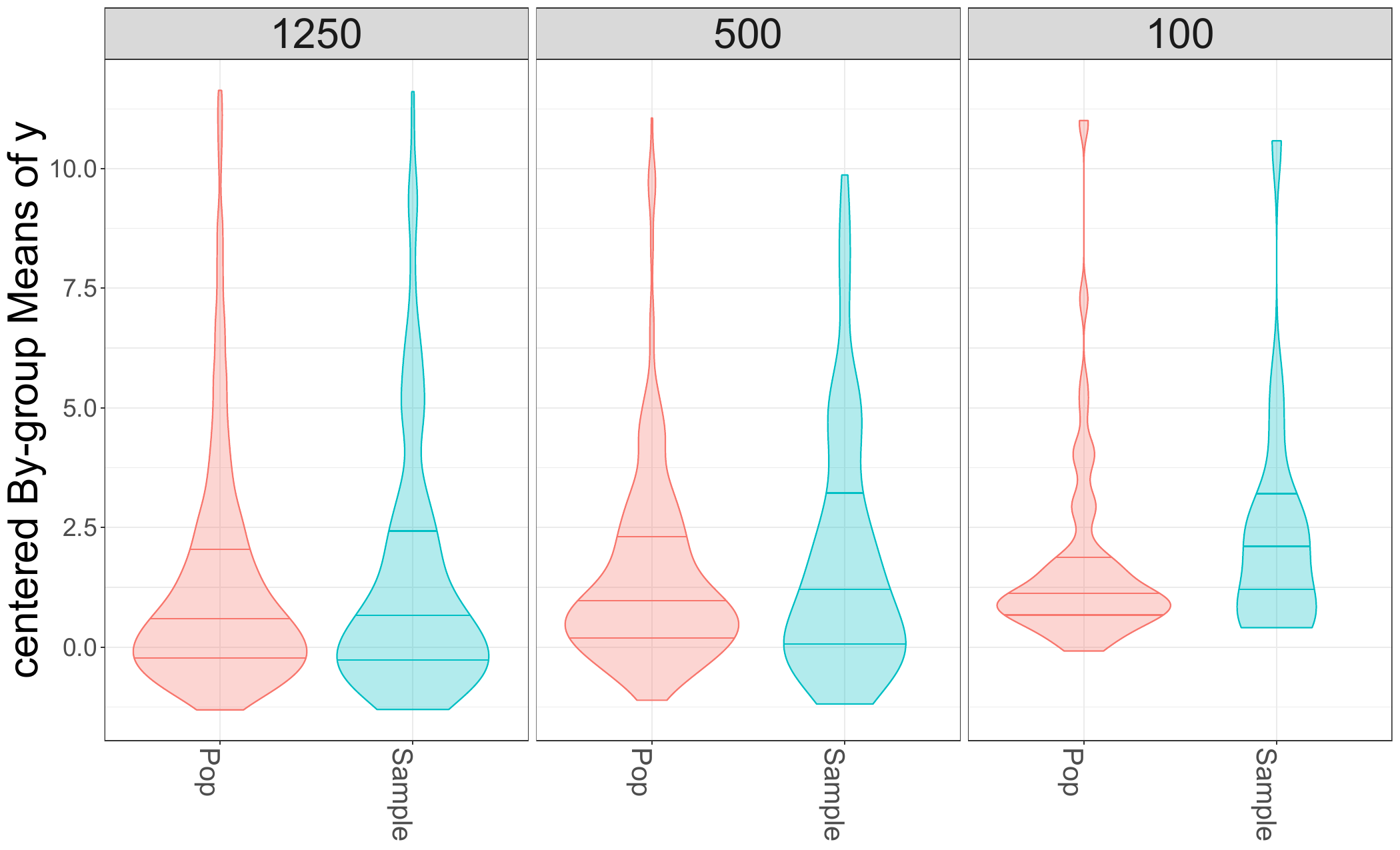}
\caption{Each plot panel compares distributions of $\overline{(y-\exp(x\beta))}_{g}$ for each of a synthetic Population and a single sample taken from that population, faceted by a sequence for the number of population groups, $G_{U}$. The resulting violin plots present each distribution within $95\%$ quantiles.}
\label{fig:biasgrp}
\end{figure}
\afterpage{\FloatBarrier}
To make concrete the notion of informative random effects, we generate a single population and subsequently take a single, informative sample of groups from that population of groups under a proportion-to-size design, using the procedures for population generation and the direct sampling of groups, described above.  The size for each population is $N = 5000$ and the sample size is $n = 500$.  We, next, average the item responses, $y$ in each group after centering by removing the fixed effects observed in the sample (excluding $x_{2}$).  For illustration, the computed $\left(\overline{(y-\exp(x\beta))}_{g}\right)$ will be used as a naive indicator of the distribution of the random effects, $\exp(u_{g})$. Each plot panel in Figure~\ref{fig:biasgrp} compares the distributions of this group-indexed centered mean statistic between the generated population and resulting informative sample for a population.  A collection of plot panels for a sequence of populations with $G_{U} = (1250, 500, 100)$ number of population groups is presented, from left-to-right.   Fixing a plot panel, each violin distribution plot includes horizontal lines for the $(0.25,0.5,0.75)$ quantiles.  We see that under a proportion-to-size design that the distributions for the centered, group mean statistic in the sample are different from the underlying populations and skew larger than those for the populations.  This upward skewness in each sample indicates that performing population estimation on the observed sample will induce bias for random effects variance, $\phi$, without correction of the group indexed random effects distribution in the sample, which we accomplish by weighting the distribution over random effects back to that for the population.

\subsection{Varying Number of Population Groups, $G_{U}$}\label{sim:varygroups}
\begin{figure}
\centering
\includegraphics[width = 0.80\textwidth,
		page = 1,clip = true, trim = 0.0in 0.0in 0in 0.in]{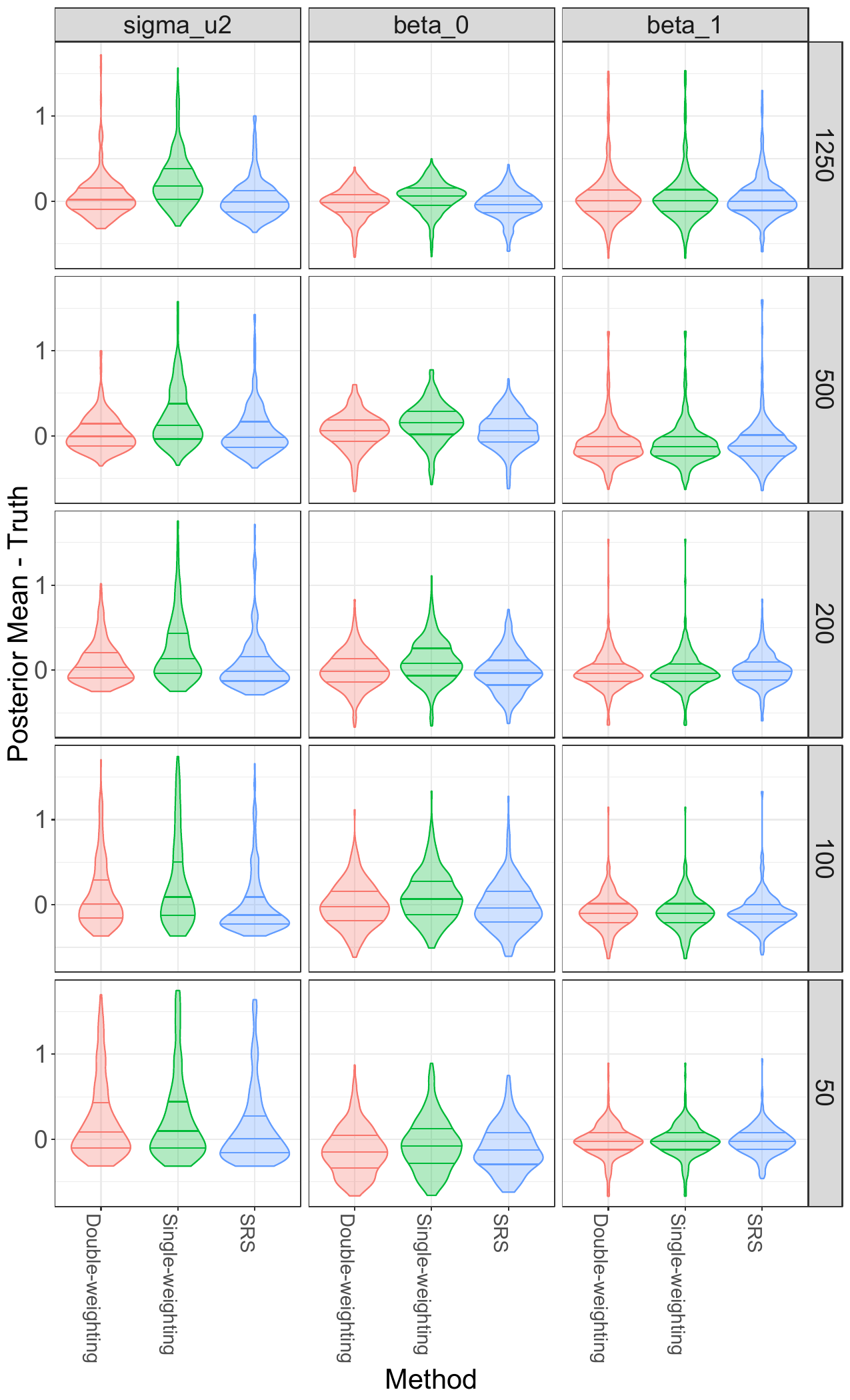}
\caption{Direct sampling of groups: Monte Carlo distributions and quantiles (0.5\%, 25\%, 50\%, 75\%, 99.5\%) for $B = 300$ iterations of differences between Posterior Means and truth under Single- and Double-weighting schema as compared to SRS for varying number of random effect groups, $G_{U}$, under $x_{2}\sim \mathcal{E}(1/2.5)$ for $N = 5000$ and $n=500$.  Parameters $(\beta_{0},\beta_{1},\sigma_{u}^{2})$ along columns and number of population groups, $G_{U}$, in descending order along the rows.}
\label{fig:groupre}
\end{figure}

We assess bias for a population model constructed using group-indexed random effects, where each group links to multiple units.   Our results presented in Figure~\ref{fig:groupre} compare our double-weighting method to single-weighting in the case we conduct a proportion-to-size \emph{direct} sampling of groups and, subsequently, sub-sample $f = 50\%$ of member units within groups.  We include an SRS sample of groups and units within selected groups taken from the same population.  The results reveal that bias is most pronounced in the case of a relatively larger number of groups e.g., $G_{U}=(1250,500)$  for $N=5000$, where each group links relatively few units.  By contrast, as the number of groups decreases, fixing the population size, $N$, the number of units linking to each group increases, which will have the effect of reducing variation in resulting sampling weights among the groups until, in the limit, there is a single group (with $\pi_{g} = 1$).  The relative bias of single-weighting, therefore, declines as the number of groups declines (and units per group increases), such that residual bias in $\sigma_{u}^{2}$ for $G_{U}=100$ is dominated by increasing variability (because we sample fewer groups) for all methods.  We, nevertheless, detect a small decrease in bias when we use double-weighting.  We include Table~\ref{tab:groupre} that presents the bias in the estimation for the posterior mean values of $(\beta_{0},\beta_{1},\sigma_{u}^{2})$ that confirms the reduction in bias for $\sigma_{u}^{2}$ under double-weighting for $G_{U} = 100$.
Our set-up may be viewed as more likely to induce bias because we assign units to groups by sorting units on the values of the size variable, $x_{2} \sim \mathcal{E}(1/(2.5))$ for allocation to groups.   Our proportion-to-size sampling design selects groups based on the mean size variable for each group, $\bar{x}_{2}$.  This set-up will tend to accentuate the variance in the resulting group-indexed size variable (and, hence, the resulting survey sample inclusion probabilities) as compared to a random allocation of units to groups.  Our simulation set-up is, nevertheless, realistic because many surveys are characterized by relatively homogenous clusters; for example, the geographically-indexed metropolitan statistical areas (MSAs) (which may be viewed as clusters) used by the Current Employment Statistics survey (administered by the Bureau of Labor Statistics) tends to express larger (higher number of employees) establishments in more highly populated MSAs.
\afterpage{\clearpage}

\rowcolors{2}{gray!6}{white}
\begin{table}[!h]

\caption{\label{tab:groupre}Difference of Single- and Double-weighting with Increasing Number of Groups, $G$, under $x_{2}\sim \mathcal{E}(1/2.5)$}
\centering
\begin{tabular}[t]{r|r|r|l|l|r}
\hiderowcolors
\toprule
$\beta_{0}$ & $\beta_{1}$ & $\sigma_{u}^{2}$ & Statistic & Method\\
\midrule
0.03 & 0.03 & 0.25 & bias & Single-weighting & 1250\\
-0.06 & 0.03 & 0.06 & bias & Double-weighting & 1250\\
-0.05 & 0.01 & 0.02 & bias & sRS & 1250\\
\hline
0.15 & -0.13 & 0.19 & bias & Single-weighting & 500\\
0.04 & -0.13 & 0.02 & bias & Double-weighting & 500\\
0.04 & -0.12 & 0.04 & bias & sRS & 500\\
\hline
0.10 & -0.02 & 0.26 & bias & Single-weighting & 200\\
-0.01 & -0.02 & 0.09 & bias & Double-weighting & 200\\
-0.02 & 0.00 & 0.05 & bias & sRS & 200\\
\hline
0.09 & -0.09 & 0.31 & bias & Single-weighting & 100\\
-0.01 & -0.10 & 0.11 & bias & Double-weighting & 100\\
-0.03 & -0.10 & 0.07 & bias & sRS & 100\\
\hline
-0.06 & -0.03 & 0.44 & bias & Single-weighting & 50\\
-0.14 & -0.03 & 0.29 & bias & Double-weighting & 50\\
-0.13 & -0.02 & 0.27 & bias & sRS & 50\\
\bottomrule
\end{tabular}
\end{table}
\rowcolors{2}{white}{white}
\afterpage{\FloatBarrier}

We next compare our double-weighting approach to the best available method in the literature, the pairwise composite likelihood method of \citet{yi:2016}, specified in Equation~\ref{composite} , which we refer to as ``pair-integrated''. We compare both methods in the case of relatively few units linked to each group (e.g., $G = 500, 1250$) because \citet{yi:2016} demonstrate superior bias removal properties as compared to \citet{rh:1998} in this setup.  We exclude smaller values of $G$ because as the number of individuals within each group grows, the number of pairwise terms to include in the pair-integrated method grows quadratically.  Our simulation set-up conducts a first-stage proportion to size sampling of groups in exactly the same manner as the previous simulation study.  We additionally include an SRS of groups and, in turn, units within groups, as a benchmark. The custom R code to implement the ``pair-integrated" point estimation can be found in Appendix \ref{sec:pairwisecode}.

Figure \ref{fig:groupre_pair} presents the Monte Carlo distributions for parameter estimates, where the columns denote parameters, $(\beta_{0},\beta_{1},\sigma_{u}^{2})$, and the rows denote number of population groups, $G_{U}$.  The results demonstrate that double-weighting leads to unbiased estimation of both the fixed effects parameters and the random effects variance relative to using a two-stage SRS sample. By contrast, the pair-integrated method demonstrates both bias and variability for the random effects variance, $\sigma_{u}^{2}$, which is exactly the set-up where it is hoped to perform relatively well. This bias for pair-integrated in the random effects variance also induces bias for the fixed effects intercept, $\beta_{0}$.  As mentioned in Section~\ref{sec:direct} the pair-integrated method integrates out the random effects (from the unweighted prior distribution) \emph{before} applying the group weights, which fails to fully correct for the informative sampling of groups.  Our method, by contrast, weights the prior the random effects and integrates them out numerically through our MCMC (after sampling the random effects).  It bears mention that \cite{yi:2016} only evaluate informative sampling of units within groups, but not the informative sampling of the groups themselves, which may be why the estimation bias for $\sigma_{u}$ was not discovered.
\begin{figure}
\centering
\includegraphics[width = 0.8\textwidth,
		page = 1,clip = true, trim = 0.0in 0.0in 0in 0.in]{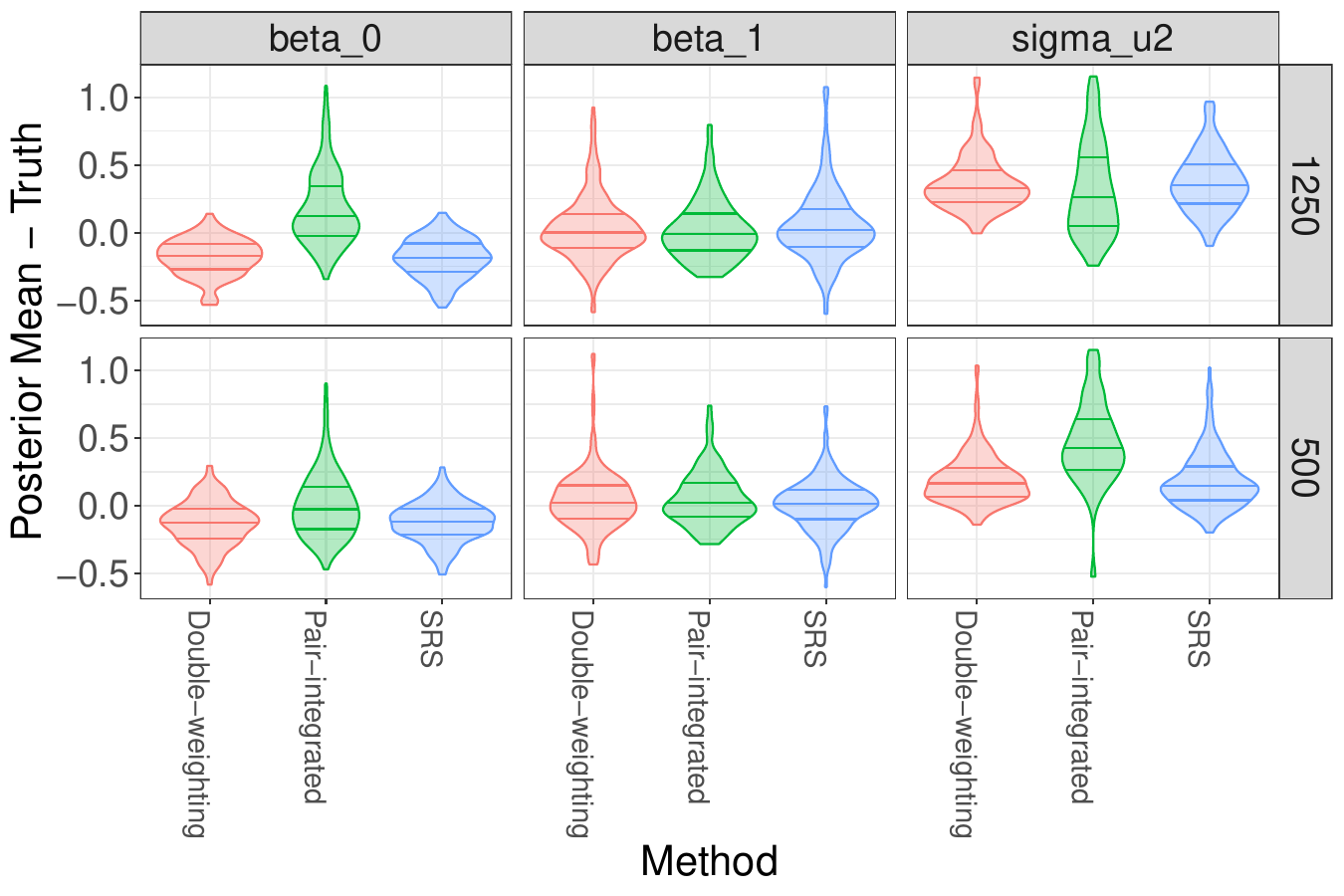}
\caption{Direct sampling of groups: Monte Carlo distributions and quantiles (0.5\%, 25\%, 50\%, 75\%, 99.5\%) for $B = 300$ iterations within a $99\%$ interval for difference of Posterior Means and truth under Double-weighting,  Pair-integrated estimation and Simple Random Sampling (SRS). Parameters $(\beta_{0},\beta_{1},\sigma_{u}^{2})$ (columns); Varying number of random effects, $G_{U}$ (rows), under $x_{2}\sim \mathcal{E}(1/2.5)$, for a population of $N = 5000$.}
\label{fig:groupre_pair}
\end{figure}
\afterpage{\clearpage}

We briefly comment on the simulation study for the indirect sampling of of groups detailed in Appendix \ref{sec:simstudy}. The results accord with the direct sampling of groups where double-weighting outperforms single-weighting.  When the number of population groups, $G_{U}$, is small, however, noise induced by sampling error results in double-weighting under-performing compared to SRS.  Yet, as the number of units per group increases with $G_{U} = 500$, the sum-weights approach outperforms SRS, which is expected because the pps design is generally more efficient such that the contraction rate of the estimator on the truth will be faster for pps (occur at a lower sample size).

Lastly, we note that while we have focused on a simple Poisson random effects formulation, our survey-weighted pseudo Bayesian posterior method readily extends to any number of levels and simultaneous employment of multiple sets of random effects without any modification to the approach. Competitor methods, by contrast, are not readily estimable.

\section{Application}\label{application}
We compare single- and double-weighting under a linear mixed effects model estimated on a dataset published by the Job Openings and Labor Turnover Survey (JOLTS), which is administered by BLS on a monthly basis to a randomly-selected sample from a frame composed of non-agricultural U.S. private (business) and public establishments.  JOLTS focuses on the demand side of U.S. labor force dynamics and measures job hires, separations (e.g. quits, layoffs and discharges) and openings.  We construct a univariate count data population estimation model with our response, $y$, defined to be the number of hires.  We formulate the associated log mean with,
\begin{equation}
\log~\mu_{i} = \mathbf{x}_{i}^{'}\bm{\beta} + u_{g\{i\}},
\end{equation}
where groups, $g = 1,\ldots,(G=892)$, denote industry groupings (defined as $6-$ digit North American Industry Classification (NAICS) codes) that collect the participating business establishments.  We expect a within-industry dependence among the hiring levels for business establishments since there are common, industry-driven economic factors that impact member establishments.  We construct the fixed effects predictors, $\mathbf{x} = \left[1,\text{ownership status},\text{region}\right]$, which are categorical predictors where ownership status holds four levels, $1.$ Private; $2.$ Federal government; $3.$ State government; $4.$ Local government.  The region predictor holds four levels, $1.$ Northeast; $2.$ South; $3.$ Midwest; $4.$ West.  Private and Northeast are designated as the reference levels.

The JOLTS sampling design assigns inclusion probabilities (under sampling without replacement) to establishments to be proportional to the number of employees for each establishment (as obtained from the Quarterly Census of Employment and Wages (QCEW)).  This design is informative in that the number of employees for an establishment will generally be correlated with the number of hires, separations and openings.  We perform our modeling analysis on a May, $2012$ data set of $n = 9743$ responding establishments.
We \emph{a priori} expect the random effects, $(u_{g})$, to be informative since larger-sized establishments would be expected to express larger variances in their hiring levels.   We choose the sum-weights method for inducing industry-level weights (from Equation~\ref{eq:sampindirmodel}) to construct our double-weighted estimation model on the observed sample.
\begin{figure}
\centering
\includegraphics[width = 0.70\textwidth,
		page = 1,clip = true, trim = 0.0in 0.0in 0in 0.0in]{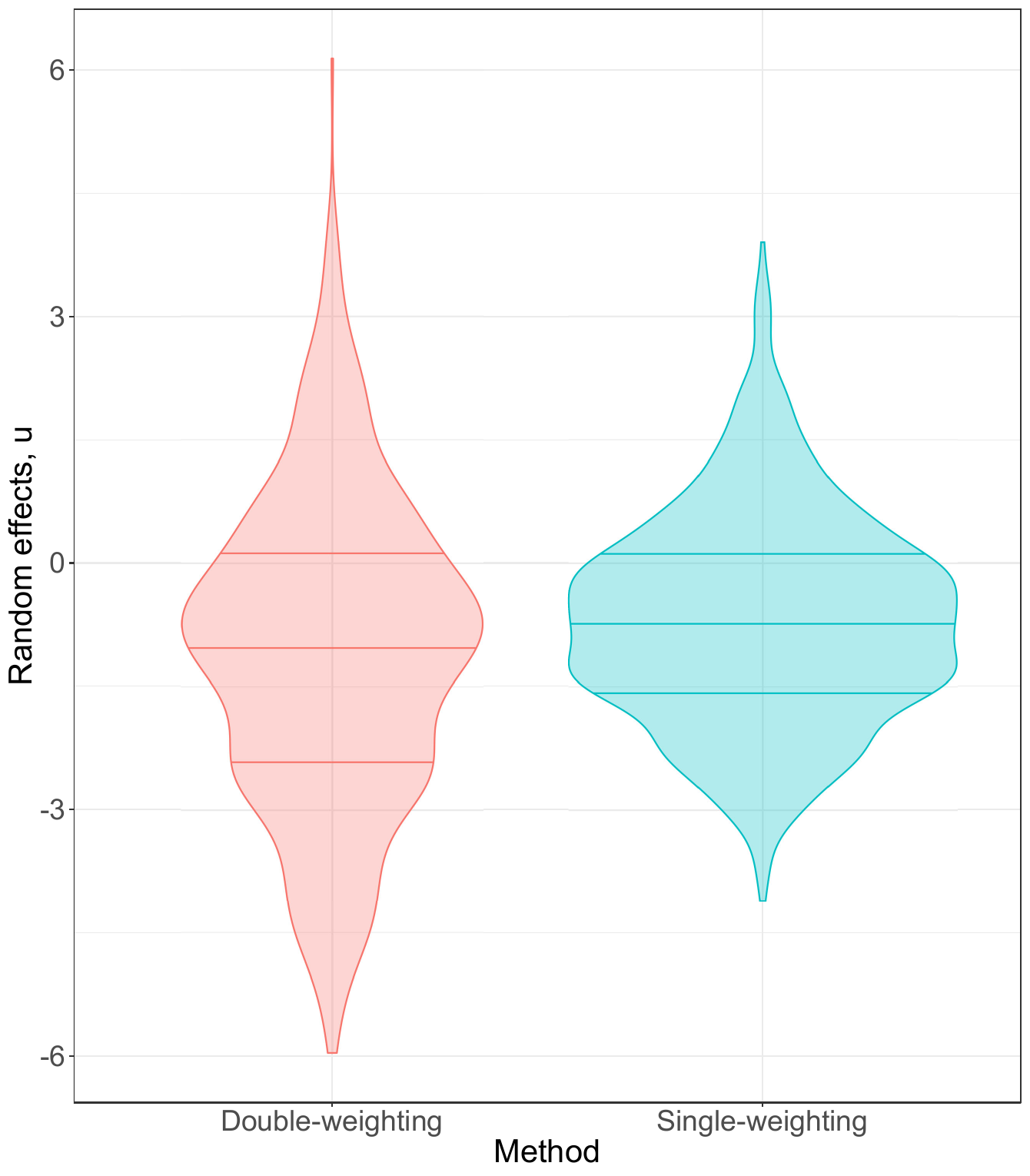}
\caption{Distributions and quantiles (25\%, 50\%, 75\%) of estimated posterior mean values for random effects, $u_{g},~ g = 1,\ldots,(G=892)$, for the JOLTS application under Single- and Double-weighting.}
\label{fig:joltsu}
\end{figure}
\afterpage{\FloatBarrier}

\begin{figure}
\centering
\includegraphics[width = 0.35\textwidth,
		page = 1,clip = true, trim = 0.0in 0.0in 0in 0.0in]{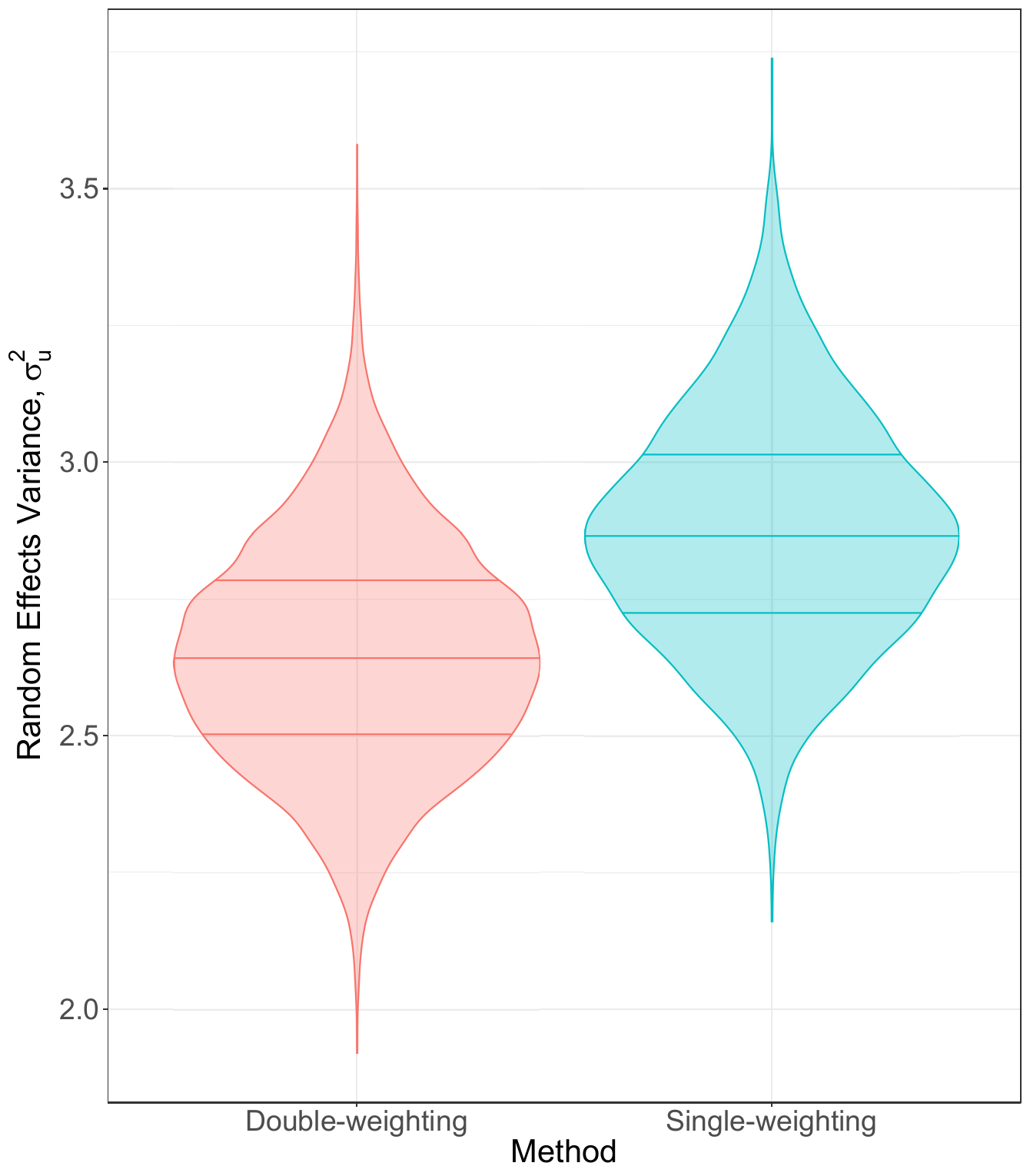}
\includegraphics[width = 0.35\textwidth,
		page = 1,clip = true, trim = 0.0in 0.0in 0in 0.0in]{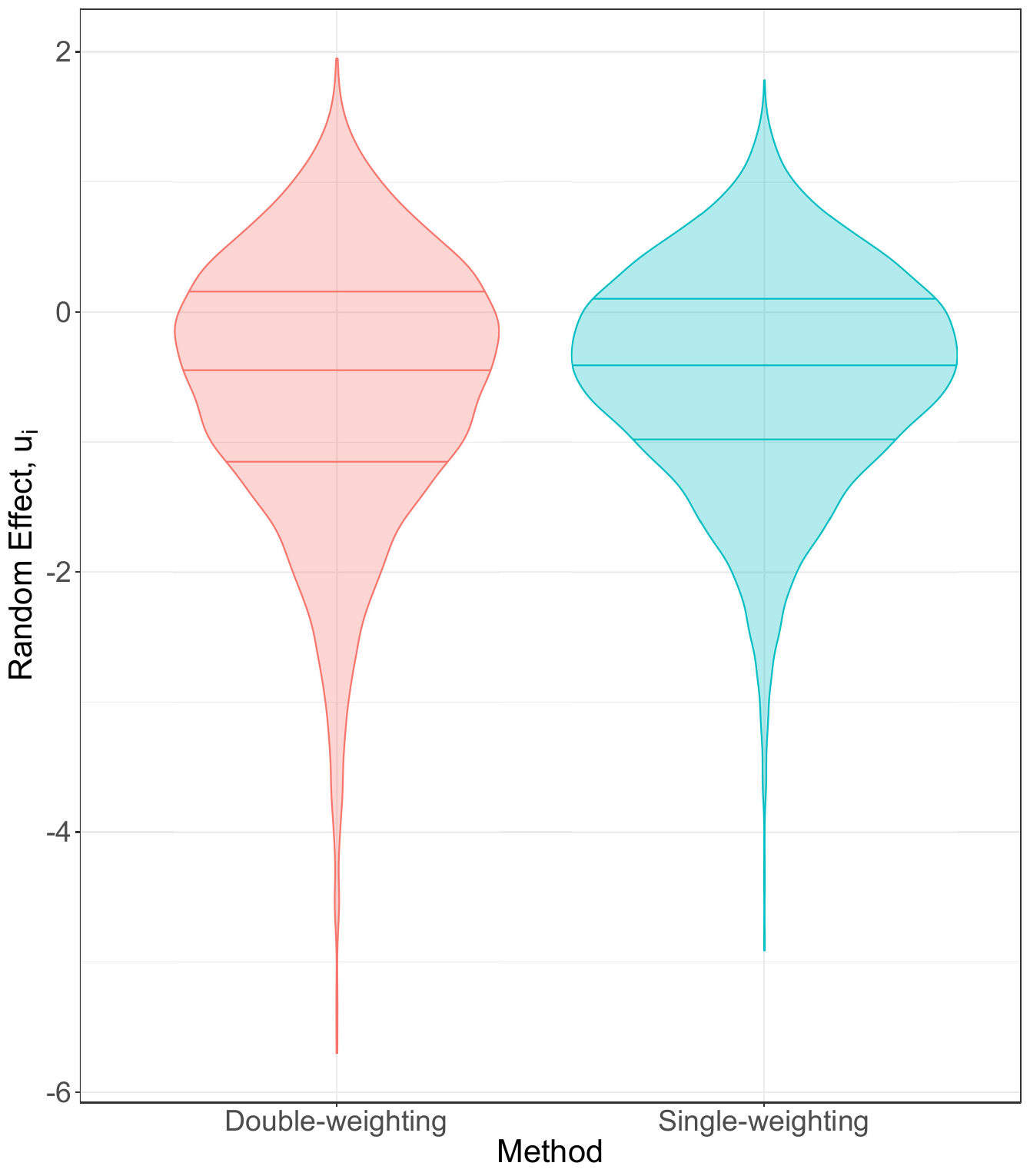}
\caption{Distributions and quantiles (25\%, 50\%, 75\%) of posterior samples for $\sigma_{u}^{2}$, the generating variance for random effects, and a single random effect parameter, $u_{i}$,  for the JOLTS application, under Single- and Double-weighting.}
\label{fig:joltsig}
\end{figure}
\afterpage{\FloatBarrier}

The more diffuse distribution over the $G = 892$ posterior mean values for random effects, $(u_{g})$, under double-weighting than single-weighting shown in Figure~\ref{fig:joltsu} demonstrates that co-weighting the likelihood and random effects distribution produces notably different inference for the group-indexed random effects; in particular, the observed sample is more homogenous in the number of hires by setting inclusion probabilities to concentrate or over-sample large-sized establishments relative to the more size-diverse population of establishments.  So the weighting of the random effects distributions in the observed sample produces a distribution over the posterior mean values for the random effects that better reflects the size-diversity of establishments in the underlying population from which the sample was taken.  Figure~\ref{fig:joltsig} presents the estimated pseudo-posterior distributions for the generating random effects variance, $\sigma_{u}^{2}$ and also a single random effect parameter, $u_{i}$, under both single- and double-weighting.  This figure reinforces the observed result for the random effects where the observed hiring levels in the survey data are more homogenous than those in the underlying population, which induces a larger posterior variation in the estimated random effects parameters for double-weighting.

\section{Discussion}\label{discussion}
In this work, we demonstrate the existence of biased estimation of both fixed \emph{and} random effects parameters when performing inference on data generated from a complex survey sample. This risk is largely unrecognized in the Bayesian literature. The current remedies come from the survey literature and are motivated from a frequentist perspective. They provide an incomplete and somewhat ad-hoc approach to the solution.  We present a principled development of the ``double-weighting'' approach based on the joint distribution of the population generating model of inferential interest and the complex sampling design represented by sample inclusion indicators. We exploit the latent variable formulation of mixed models and their related posterior sampling techniques to avoid awkward numerical integration required for frequentist solutions. We show that this simplicity also leads to reductions in bias.

This work culminates recent developmental work in combining traditional survey estimation approaches with the Bayesian modeling paradigm. The pseudo-posterior framework simultaneously offers complex survey data analysis to Bayesian modelers and the full suite of hierarchical Bayesian methods to those well-versed in traditional fixed effect analysis of survey data.

\appendix
\section{Alternative Pseudo Group Inclusion Probabilities Under Indirect Sampling}\label{sec:pseudoprobs}
If we invert the resultant group-indexed weight, $w_{g}= 1/N_{g}\times\mathop{\sum}_{j \in S_{g}}w_{g j}$, from Equation~\ref{eq:sampindirmodel}, where groups are not directly sampled, we may view the inverse of the group $g$ weight, $\tilde{\pi}_{g} = 1/w_{g}$, as a ``pseudo" group inclusion probability, since we don't directly sample groups.  The construction for one form of $\tilde{\pi}_{g}$ motivates our consideration of other formulations for the pseudo group inclusion probabilities that we may, in turn, invert to formulate alternative group-indexed survey sampling weights, $(w_{g})$.

The resulting $w_{g}$ of Equation~\ref{eq:sampindirmodel} requires either knowledge of $N_{g}$ or a method for its approximation.  The sum of nested unit weights is further composed as a harmonic sum of inverse inclusion probabilities of member units in each group, which may be overly dominated by units with small unit inclusion probabilities.  Our first alternative more directly constructs a pseudo group inclusion probability as the union of probabilities for inclusion of any member unit in the observed sample (in which case the group will be represented in the sample) and does not require estimation of population quantities, such as $N_{g}$.  Under a weak assumption of nearly independent sampling within groups, this alternative is constructed as,
\begin{align}
\tilde{\pi}_{g} &= \mathop{\sum}_{\ell = 1}^{N_{g}}\pi_{\ell}\label{eq:sprpop}\\
\hat{\tilde{\pi}}_{g} &= \mathop{\sum}_{\ell = 1}^{N_{g}}\frac{\delta_{\ell}}{\pi_{\ell}} \times \pi_{\ell}\\
 &= \mathop{\sum}_{j=1}^{n_{g}} w_{j} \times \pi_{j}\label{eq:sprsamp}
\end{align}
where $\pi_{\ell}$ denotes the marginal inclusion probability for unit, $\ell \in (1,\ldots,N_{g})$, where we recall that $N_{g}$ denotes the number of units linked to group, $g \in(1,\ldots,G_{U})$, in the population of groups.  We may estimate the pseudo group inclusion probabilities in the observed sample by making the same walk from population-to-observed-sample as is done in Equation~\ref{estimator} to Equation~\ref{estimatorsample}; by including unit sampling weights, $(w_{j})_{j \in S_{g}}$ ($S_{g} = \{1,\ldots,n_{g}\}$).  We normalize the $(w_{j})_{j \in S_{g}}$ to sum to $1$ as our intent is to re-balance the information (among sampled units) within a group to approximate that of the population of units within the group.  While this estimator has the undesirable property of computing $\tilde{\pi}_{g} > 1$, we utilize this quantity to weight the random effects prior density contributions with, $w_{g} \propto 1/\tilde{\pi}_{g}$, so we focus on the effectiveness of estimation bias removal for generating hyperparameters of the $(u_{g})_{g \in G_{U}}$.  We label this method as the ``sum-probabilities" method in contrast to the ``sum-weights" methods with which we label the result of Equation~\ref{eq:sampindirmodel}.

Our second alternative for estimation of a pseudo group inclusion probability is designed to ensure $\tilde{\pi}_{g} \leq 1$ by using a product complement approach that computes the union of member unit probabilities for a group, indirectly, by first computing its complement and subtracting that from $1$.  To construct this estimator, we assume that units, $j \in S$ are sampled independently \emph{with} replacement, which is a tenable assumption when drawing a small sample from a large population of units.  Let $\pi^{(1)}_{j}$ denote the probability of selecting unit, $j$, in a sample of size $1$ (e.g., a single draw).   Then we may construct the marginal inclusion probability of unit, $\pi_{j}$, for a sample of size, $n = \vert S\vert$, as the complement that unit $j$ does not appear in any of the $n$ draws,
\begin{equation}\label{eq:pwr}
\pi_{j} = 1- \left(1-\pi^{(1)}_{j}\right)^{n},
\end{equation}
where $\mathop{\sum}_{j\in U}\pi^{(1)}_{j} = 1$.   By extension, $0 < \tilde{\pi}^{(1)}_{g} = \mathop{\sum}_{j \in U_{g}}\pi^{(1)}_{j} \leq 1$, where $\tilde{\pi}^{(1)}_{g}$ denotes the pseudo group, $g \in (1,\ldots,G_{U})$ inclusion probability for a sample of size $1$ and is composed as the union of size $1$ probabilities for member units.  The expression for the pseudo group inclusion probability derives from the underlying sampling of members with replacement,
\begin{equation}\label{eq:gpwr}
\tilde{\pi}_{g} = 1 - \left(1-\tilde{\pi}^{(1)}_{g}\right)^{n} = 1 - \left(1-\mathop{\sum}_{j=1}^{N_{g}}\pi^{(1)}_{j}\right)^{n},
\end{equation}
where we exponentiate the complement term, $\left(1-\tilde{\pi}^{(1)}_{g}\right)$, by the number of draws of units, $n$, (rather than $G_{S}$, the number of groups represented in the observed sample) because we don't directly sample groups.  We solve for $\pi^{(1)}_{j}$ using Equation~\ref{eq:pwr}, $\pi^{(1)}_{j} = 1 - \left(1-\pi_{j}\right)^{(1/n)}$, and plug into Equation~\ref{eq:gpwr} to achieve,
\begin{align}
\tilde{\pi}_{g} &= 1 - \left(1-\mathop{\sum}_{j=1}^{N_{g}}\left(1-\left(1-\pi_{j}\right)^{(1/n)}\right)\right)^{n}\\
\hat{\tilde{\pi}}_{g} &= 1 - \left(1-\mathop{\sum}_{j=1}^{N_{g}}\frac{\delta_{j}}{\pi_{j}}\left(1-\left(1-\pi_{j}\right)^{(1/n)}\right)\right)^{n}\\
&= 1 - \left(1-\mathop{\sum}_{\ell=1}^{n_{g}}w_{\ell}\left(1-\left(1-\pi_{\ell}\right)^{(1/n)}\right)\right)^{n}\label{eq:pcprsamp},
\end{align}
where, as with the sum-probabilities formulation, we normalize the unit weights within each group, $(w_{\ell})_{\ell \in S_{g}}$, to sum to $1$.  We label this method as ``product-complement".

\section{Simulation Study Results for Alternative Pseudo Group Inclusion Probabilities}\label{sec:simstudy}
We present the results for the simulation study that samples units, rather than groups, for the expanded set of methods developed in Appendix \ref{sec:pseudoprobs} for the pseudo group inclusion probabilities.  We recall that under this single stage sampling of units, groups are not directly sampled under the survey sampling and are included to the extent that one or more member units are sampled. 

The synthetic population (for each Monte Carlo iteration) utilizes group-indexed random effects under size-based assignment of population units to groups under each alternative for total number of groups, $G_{U}$.  In this study, we randomly vary the number of population units assigned to each group with the mean values for each $G_{U}$ set to be equal to the fixed number of units per group.  We allocate a relatively higher number of units to those groups with smaller-sized units under each group size, $G_{U}$, to mimic our application. The number of population units per group, $N_{h}$, is set to randomly vary among the $G_{U}$ population groups using a log-normal distribution centered on the $(4, 10, 25, 50, 100)$ units per group used in the case of direct sampling, with a variance of $0.5$.  In the case of $G_{U} = 1250$, this produces a right-skewed distribution of the number of units in each group, ranging from approximately $1$ to $40$ units per group and the total number of population units per group is restricted to sum to $N = 5000$.

We sort the groups such that groups with larger-sized units are assigned relatively fewer units and groups with smaller-sized units are assigned relatively more units.   This set-up of assigning more units to smaller-sized groups mimics the estimation of employment counts among business establishments analyzed in our application in the sequel, where there are relatively few establishments with a large number of employees (e.g., $> 50$) (which is the size variable), while there are, by contrast, many more establishments (small businesses) that have a small number of employees (e.g., $< 10$).

The survey sampling design employed here is a \emph{single-stage}, proportion-to-size design that directly samples the units (not the groups) with unit inclusion probabilities proportional to the size variable, $x_{2} \sim \mathcal{E}(1/(m_{2}=3.5))$.
Each Monte Carlo iteration of our simulator (that we run for $B = 300$ iterations) generates the population $(y_{i},x_{1i},x_{2i})_{i=1}^{N}$, assigns group and unit inclusion probabilities for the population in the case of direct sampling of groups or assigns unit inclusion probabilities in the case of indirect sampling.  A sample of $n = 500$ is then taken and estimation is performed for  $(\beta_{0},\beta_{1},\sigma_{u}^{2})$ from the observed sample under three alternatives:
\begin{enumerate}
\item Single-weighting, where we solely exponentiate the likelihood contributions for $(y_{g j})$ by sampling weights, $(w_{g j} \propto 1/\pi_{g j})$ (and don't weight the prior for the random effects, $(u_{g})$);
\item Double-weighting, where we exponentiate \emph{both} the likelihood for the $(y_{g j})$ by sampling weights, $(w_{g j})$, and also exponentiate the prior distribution for $u_{g}$ by  weight, $w_{g} \propto 1/\pi_{g}$ (for each of $g = 1,\ldots,G_{S}$).  We estimate $\pi_{g}$ using each of the three methods presented in Appendix \ref{sec:pseudoprobs}: ``sum-weights'', ``sum-probabilities'', and ``product-complement''.
\item SRS, where we take a single-stage simple random sample of units.
\end{enumerate}
The inclusion of model estimation under (a non-informative) SRS is intended to serve as a gold standard against which we may judge the bias and MSE performance of single- and double-weighting under informative sampling.

\begin{figure}
\centering
\includegraphics[width = 0.90\textwidth,
		page = 1,clip = true, trim = 0.0in 0.0in 0in 0.in]{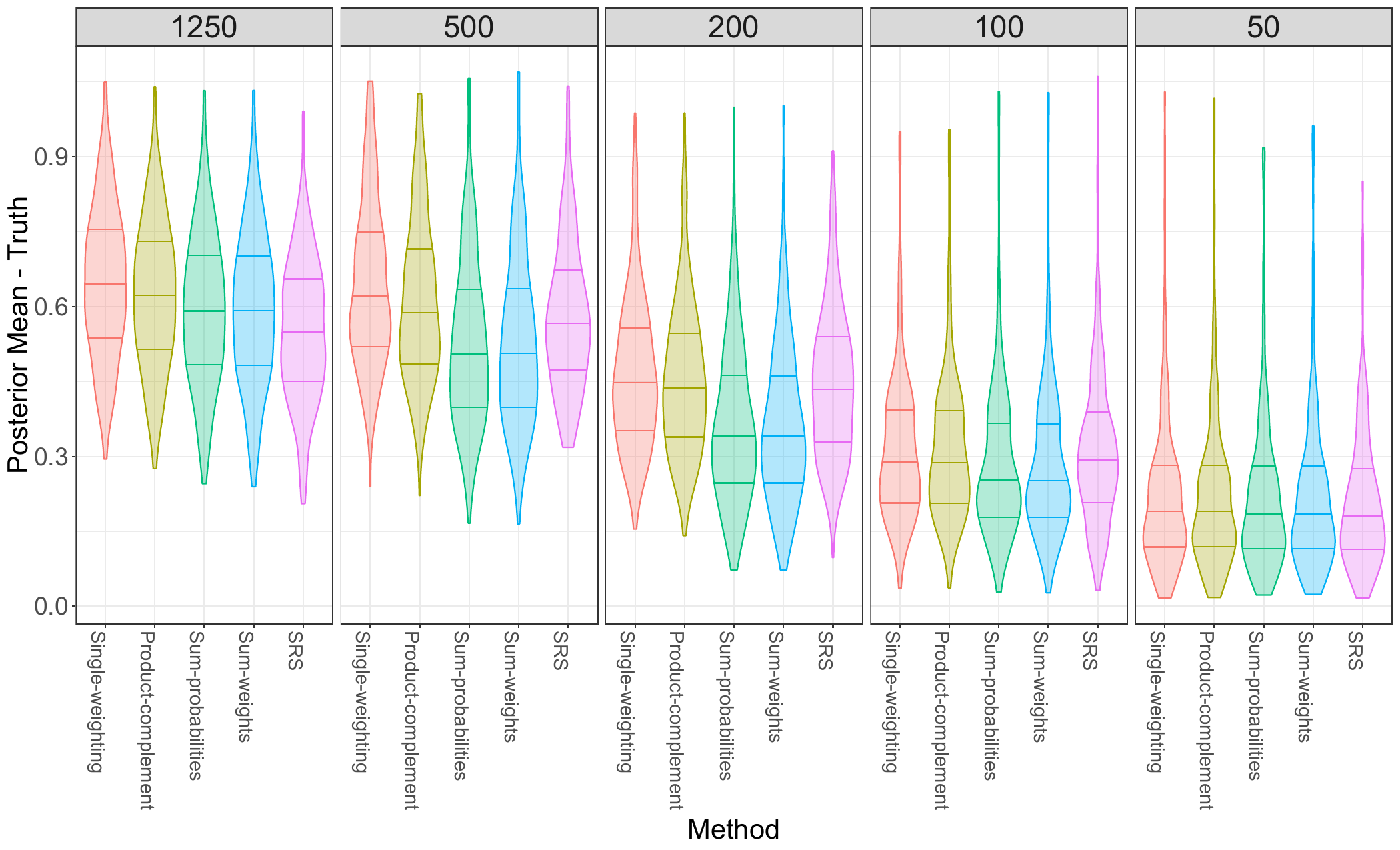}
\caption{Indirect sampling of groups: Monte Carlo distributions and quantiles (0.5\%, 25\%, 50\%, 75\%, 99.5\%) for $B = 300$ iterations for $\sigma_{u}^{2}$ for difference of posterior means and truth under alternative weighting schema for varying number of groups, $G$. Population of $N = 5000$ and sample of $n = 500$.  In each plot panel, from left-to-right is the Single-weighting method, Product Complement double-weighting method of Equation~\ref{eq:pcprsamp}, Sum-probabilities double-weighting method of Equation~\ref{eq:sprsamp}, Sum-weights double-weighting method of Equation~\ref{eq:sampindirmodel} and Simple-random sampling (SRS).}
\label{fig:inducegrprealts}
\end{figure}

Each plot panel in Figure~\ref{fig:inducegrprealts} shows the distributions over Monte Carlo simulations for estimates of the generating variance, $\sigma_{u}^{2}$, of the random effects, $(u_{g})$, under each of the following weighting methods: single-weighting, product-complement double-weighting (Equation~\ref{eq:pcprsamp}), sum-probabilities double-weighting (Equation~\ref{eq:sprsamp}), sum-weights double-weighting (Equation~\ref{eq:sampindirmodel}), and SRS (no weighting under simple random sampling of the same population from which the pps sample was taken).  The panels are ordered from left-to-right for a sequence of $G_{U} = \left(1250,500,200,100,50\right)$.  As earlier mentioned, the number of population units per group, $N_{h}$, is set to randomly vary under a lognormal distribution, though there will on average be more units sampled per group from synthetic populations with a smaller number of population groups, $G_{U}$, than there will be units per group sampled under a larger number of population groups.  The sum-probabilities and sum-weights methods for accomplishing double-weighting generally perform nearly identically to one another and better than single-weighting for all group sizes.  Since sum-probabilities and sum-weights perform nearly identically, one may choose to prefer use of the former because it does not require our estimation of $\hat{N}_{g}$, as does the latter.

Table~\ref{tab:inducegrprealts} presents the relative bias, defined as the bias divided by the true value, and the normalized root MSE, defined as the square root of MSE divided by the true value, for the regression coefficients, $(\beta_{0},\beta_{1})$, to accompany Figure~\ref{fig:inducegrprealts}.  We show the relative bias and normalized RMSE quantities in this study because the true values of the marginal model, $\sigma_{u}^{2} = \left(0.578,0.349,0.216,0.169,0.136\right)$, varies over the sequence of sizes for $G_{U}$.  As in the case of direct sampling of groups, there is an association between the amount bias in estimation of $\sigma_{u}^{2}$ and in the intercept coefficient, $\beta_{0}$.
\afterpage{\FloatBarrier}

\rowcolors{2}{gray!6}{white}
\begin{table}[!h]
\caption{\label{tab:inducegrprealts} Normalized Bias and RMSE for Double-weighting methods as compared to Single-weighting and SRS for Increasing Units Per Random Effect Under \emph{Indirect} Sampling of Groups for $B = 300$ iterations with population of $N = 5000$ and sample of $n = 500$.}
\centering
\begin{adjustbox}{width=0.6\textwidth}
\begin{tabular}{l|r|r|r|r|r}
\toprule
\multicolumn{2}{c}{ } & \multicolumn{2}{c}{Relative Bias} & \multicolumn{2}{c}{Normalized RMSE} \\
\cmidrule(l{3pt}r{3pt}){3-4} \cmidrule(l{3pt}r{3pt}){5-6}
Model & G & beta\_0 & beta\_1 & beta\_0 & beta\_1\\
\midrule
\rowcolor{gray!6}  Product-complement & 1250 & -0.54 & -0.01 & 0.58 & 0.11\\
Single-weighting & 1250 & -0.55 & 0.00 & 0.58 & 0.11\\
\rowcolor{gray!6}  SRS & 1250 & -0.58 & 0.00 & 0.61 & 0.11\\
Sum-probabilities & 1250 & -0.52 & -0.01 & 0.55 & 0.11\\
\rowcolor{gray!6}  Sum-weights & 1250 & -0.52 & -0.01 & 0.55 & 0.11\\
\addlinespace
Product-complement & 500 & -0.45 & 0.01 & 0.48 & 0.13\\
\rowcolor{gray!6}  Single-weighting & 500 & -0.45 & 0.01 & 0.48 & 0.14\\
SRS & 500 & -0.46 & 0.00 & 0.49 & 0.11\\
\rowcolor{gray!6}  Sum-probabilities & 500 & -0.41 & 0.00 & 0.45 & 0.13\\
Sum-weights & 500 & -0.41 & 0.00 & 0.45 & 0.13\\
\addlinespace
\rowcolor{gray!6}  Product-complement & 200 & -0.26 & 0.00 & 0.34 & 0.12\\
Single-weighting & 200 & -0.26 & 0.00 & 0.34 & 0.12\\
\rowcolor{gray!6}  SRS & 200 & -0.29 & 0.00 & 0.36 & 0.11\\
Sum-probabilities & 200 & -0.23 & 0.00 & 0.33 & 0.12\\
\rowcolor{gray!6}  Sum-weights & 200 & -0.23 & 0.00 & 0.33 & 0.12\\
\addlinespace
Product-complement & 100 & -0.13 & -0.02 & 0.30 & 0.13\\
\rowcolor{gray!6}  Single-weighting & 100 & -0.13 & -0.02 & 0.30 & 0.13\\
SRS & 100 & -0.19 & 0.00 & 0.33 & 0.12\\
\rowcolor{gray!6}  Sum-probabilities & 100 & -0.13 & -0.02 & 0.30 & 0.13\\
Sum-weights & 100 & -0.13 & -0.02 & 0.30 & 0.13\\
\addlinespace
\rowcolor{gray!6}  Product-complement & 50 & -0.07 & 0.00 & 0.39 & 0.11\\
Single-weighting & 50 & -0.07 & 0.00 & 0.39 & 0.11\\
\rowcolor{gray!6}  SRS & 50 & -0.08 & -0.01 & 0.39 & 0.11\\
Sum-probabilities & 50 & -0.07 & 0.00 & 0.39 & 0.11\\
\rowcolor{gray!6}  Sum-weights & 50 & -0.07 & 0.00 & 0.39 & 0.11\\
\bottomrule
\end{tabular}
\end{adjustbox}
\end{table}
\rowcolors{2}{white}{white}
\afterpage{\FloatBarrier}

\section{Stan Code for Estimating Poisson Mixed Effects Model }\label{sec:stanscript}

We next present the Stan \citep{stan:2015} script that enumerates the probability model specification of the Poisson likelihood and associated prior distributions.  We utilized Stan to estimate the mixed effects Poisson model implemented in Sections~\ref{simulation} and \ref{application}.

{\small
\begin{lstlisting}[language=R]
functions{

  real wt_pois_lpmf(int[] y, vector mu, vector weights, int n){
    real check_term;
    check_term  = 0.0;
    for( i in 1:n )
    {
	  check_term    = check_term + weights[i] * poisson_log_lpmf(y[i] | mu[i]);
    }
    return check_term;
  }
  
  real normalwt_lpdf(vector y, vector mu, real sigma, vector weights, int n)
  {
    real check_term;
    check_term  = 0.0;
    for( i in 1:n )
    {
      check_term    = check_term + weights[i] * normal_lpdf(y[i] | mu[i], sigma);
    }
    return check_term;
  }
  
} /* end function{} block */



data {
    int<lower=1> n; // number of observations
	  int<lower=1> K; // number of linear predictors
	  int<lower=1> G; // number of random effects groups
	  int<lower=0> s_re[n]; // assignment of groups to items
    int<lower=0> y[n]; // Response variable
    vector<lower=0>[n] weights; // observation-indexed (sampling) weights
    vector<lower=0>[G] weights_re; // group-indexed (sampling) weights
    matrix[n, K] X; // coefficient matrix
}

transformed data{
  vector<lower=0>[G] zeros_g;
  vector<lower=0>[K] zeros_beta;
  zeros_beta  = rep_vector(0,K);
  zeros_g  = rep_vector(0,G);
} /* end transformed data block */

parameters{
  vector[K] beta; /* regression coefficients from linear predictor */
  vector<lower=0>[K] sigma_beta;
  /* cholesky of correlation matrix for Sigma_beta */
  cholesky_factor_corr[K] L_beta; 
  vector[G] u;
  real<lower=0> sigma_u;
}

transformed parameters{
  real<lower=0> sigma_u2;
  vector[n] mu;
  vector[n] fixed_effects;
  fixed_effects   = X * beta;
  mu      = fixed_effects + u[s_re];
  sigma_u2        = pow(sigma_u,2);
} /* end transformed parameters block */

model{
  L_beta          ~ lkj_corr_cholesky(6);
  sigma_u         ~ student_t(3,0,1);
  sigma_beta      ~ student_t(3,0,1);
  /* Implement a beta ~ N_{T}(0,Q^{-1})  */
  /* K x 1, vector */
  beta            ~ multi_normal_cholesky( zeros_beta, diag_pre_multiply(sigma_beta,L_beta) ); 
  /* directly update the log-probability for sampling */
  u               ~ normalwt(zeros_g, sigma_u, weights_re, G); // weighting the random effects
  target          += wt_pois_lpmf(y | mu, weights, n);
} /* end model{} block */

\end{lstlisting}
}

\section{R Code for Pairwise Integrated Poisson Mixed Effects Model }\label{sec:pairwisecode}

We next present the R \citep{R} scripts to implement the pairwise integrated likelihood approach of \citep{yi:2016} for Poison models.  We utilized this model in Sections~\ref{simulation}.

{\small
\begin{lstlisting}[language=R]
library(fastGHQuad)

##Poison model integrated likelihood
#joint density - does not include exp(-u^2) which is included in ghQuad
#change of variables z = sqrt(2)*u*sig.z,
#where z ~N(0, sig.z) random effect

pwpois.GH <- function(u,x1,x2, mu1, mu2, sig.z){
#first argument is what is integrated out
  d <- dpois(x1, mu1*exp(sqrt(2)*u*sig.z))*
    dpois(x2, mu2*exp(sqrt(2)*u*sig.z))*1/sqrt(pi)
  return(d)
}
#see qhQuad help for more details
intpois.ghfs <- function(x1, x2, mu1, mu2, sig.z){
#scalar arguments for each
  intz <- ghQuad(pwpois.GH, rule = gaussHermiteData(5),
                 x1 = x1, x2 = x2, mu1 = mu1, mu2 = mu2, sig.z = sig.z)
  return(intz)	
}

##Weighted Logliklihood  
#assume y1 and y2 are vectors of pairs 
#with corresponding matrix predictors X1 and X2 and vector
#of pairwise weights pwt
  
wt_comp_pois_ll <- function(par, y1,y2, X1,X2, pwt){
#need vector of parameters par as input for optim
  k <- length(par)
  beta <- par[1:(k-1)]
 #have log(sig) as parameter to keep bounds -Inf to Inf in optim
  sig <- exp(par[k]) 
    
  mu1 <- exp(X1%*%beta) #need log link function for mu
  mu2 <- exp(X2%*%beta)
    
  ll <- 0
  for(i in seq_along(y1)){
    #print(i)
    inti <- intpois.ghfs(y1[i],y2[i],mu1[i], mu2[i], sig)
    ll <- ll + pwt[i]*max(log(inti), -743) #truncate to keep above -INF
    #print(ll)
  }
  return(-ll) #need to minimize -LL
}
  
## Input into optim
#Example with 3 parameters (Beta1, Beta2, Sigma_u)
## starting values
testpar   <- c(rnorm(1,0,.5), rnorm(1,0,.5), log(rgamma(1,3,rate=3))) 
est1      <- optim(par = testpar, fn = wt_comp_pois_ll, method = "BFGS", 
    y1 = y1, y2=y2, X1= X1, X2= X2, pwt = pwt)
## extract parameter estimates
pars          <- est1$par
pars[3]       <- exp(pars[3]) #back-transform to sigma in [0,Inf]
beta_hat      <- pars[1:2]
sigmau2_hat   <- pars[3]^2
\end{lstlisting}
}

\bibliography{refs_jan_2021}
\bibliographystyle{imsart-nameyear}

\end{document}